\documentclass [11pt]{article}
\usepackage{amsmath,amsthm,amsfonts,amscd,eucal,latexsym,amssymb}
\usepackage{epsfig}   
\input xy
\xyoption{all}
\def\sp{\hskip -5pt} 
\def\spa{\hskip -3pt} 

\newsymbol\bt 1202           %%% \boxtimes 

\def\cF{{\ca F}}
\def\cG{{\ca G}}

\def\cH{{\ca H}}

\def\cS{{\ca S}}
\def\cT{{\ca T}}
\def\cA{{\ca A}}
\def\cF{{\ca F}}

\def\cW{{\ca W}}

\def\sS{{\mathsf S}}

\def\bC{{\mathbb C}}           %%%  complex numbers and so on 
 
\def\bI{{\mathbb I}}

\def\bN{{\mathbb N}} 
\def\bM{{\mathbb M}} 
 
\def\bR{{\mathbb R}}

\def\bS{{\mathbb S}}

\def\bZ{{\mathbb Z}} 
 
\newsymbol \rest 1316         %%% restriction symbol 
 
\def\gB{{\mathfrak B}}

\def\gH{{\mathfrak H}} 
\def\gh{{\mathfrak h}}

\def\beq{\begin{eqnarray}}
\def\eeq{\end{eqnarray}}

\def\al{\langle}
\def\cl{\rangle}
\newcommand{\ca}[1]{{\cal #1}}         %%  calligraphic

\def\z{\zeta}
\def\bz{\overline{\zeta}}

\def\scri{\Im^+}         %null future boundary
\def\scrip{\Im^-}         %null past boundary
\def\tg{\tilde{g}}
\def\tM{\tilde{M}}

\def\tV{\tilde{V}}
\def\MV{M_{\tilde{V}}}

\def\cU{{\cal U}}
\def\bx{{\bf x}}

\newcounter{proposition}[section]
\newcounter{theorem}[section]
\newcounter{lemma}[section]
\newcounter{definition}[section]
\newcounter{remark}[section]

\def\theproposition{\thesection.\arabic{proposition}}
\def\thetheorem{\thesection.\arabic{theorem}}
\def\thelemma{\thesection.\arabic{lemma}}
\def\thedefinition{\thesection.\arabic{definition}}
\def\theremark{\thesection.\arabic{remark}}

\def\s #1 {\section{#1}}

\def\ssa #1 {\ifhmode{\par}\fi\refstepcounter{subsection}
  \noindent {\bf\thesubsection}. {\em #1}.\quad
  \addcontentsline{toc}{subsection}{\protect\numberline{\thesubsection} #1}%
  }

\def\ssb #1 {\ifhmode{\par}\fi\refstepcounter{subsection}
  \noindent {\bf\thesubsection.} {\em #1.}\quad
  \addcontentsline{toc}{subsection}{\protect\numberline{\thesubsection} #1}%
  }

\def\proposizione {\ifhmode{\par}\fi\refstepcounter{proposition}
  \noindent {\bf Proposition \theproposition}. \quad}
\def\teorema {\ifhmode{\par}\fi\refstepcounter{theorem}
  \noindent {\bf Theorem \thetheorem}. \quad}
\def\lemma {\ifhmode{\par}\fi\refstepcounter{lemma}
  \noindent {\bf Lemma \thelemma}. \quad}
\def\definizione {\ifhmode{\par}\fi\refstepcounter{definition}
  \noindent {\bf Definition \thedefinition}. \quad}
\def\remark {\ifhmode{\par}\fi\refstepcounter{remark}
  \noindent {\bf Remark \theremark}. \quad}

\begin{document} 
 
\hfill{\sl December 2005 -- Revised Version April 2006, Preprint  UTM-689} \\

%%%%%%%%%%%%%   Title %%%%%%%%%%%%%%%%%%%%%%%%%% 
 
\par 
\LARGE 
\noindent 
{\bf  Uniqueness theorem  for BMS-invariant states of scalar QFT on the null boundary of asymptotically flat
spacetimes and bulk-boundary observable algebra correspondence} \\
\par 
\normalsize 
  
%%%%%%%%%%%%%%%%%%%%%%%%%%%%%%%%%%%%%%%%%%%%% 
 
%%%%%%%%%%%% Authors %%%%%%%%%%%%%%%%%%%%%%%%%%% 

\noindent {\bf Valter Moretti\footnote{E-mail: moretti@science.unitn.it}} 

\par
\small

\noindent  
Dipartimento di Matematica, Facolt\`a di Scienze F.M.N., Universit\`a di Trento, \\
 \& Istituto Nazionale di Alta Matematica ``F.Severi'' - Unit\`a Locale di Trento,\\
 \&  Istituto Nazionale di Fisica Nucleare - Gruppo Collegato di Trento,\\
  via Sommarive 14  
I-38050 Povo (TN), Italy. \smallskip \smallskip

 \normalsize

\small 
\noindent {\bf Abstract}. {This work concerns some features of scalar QFT 
defined on the causal boundary $\scri$ of an asymptotically flat at null infinity spacetime 
and based on the BMS-invariant Weyl algebra 
$\cW(\scri)$.\\
(a) (i) It is noticed that the natural $BMS$ invariant pure quasifree state $\lambda$ on $\cW(\scri)$, recently introduced 
by Dappiaggi, Moretti an Pinamonti, enjoys positivity of the self-adjoint generator of $u$-translations
with respect to {\em every} Bondi coordinate frame $(u,\z,\bz)$ on $\scri$,
($u\in \bR$ being the affine parameter of the complete null geodesics forming $\scri$ and $\z,\bz$ complex coordinates 
on the transverse $2$-sphere). This fact may be interpreted 
as a remnant of spectral condition inherited from QFT in Minkowski spacetime (and it is the spectral condition 
 for free fields when the bulk is the very Minkowski space).
(ii) It is also proved that cluster property under $u$-displacements is valid for every (not necessarily quasifree)
 pure state on $\cW(\scri)$ which is invariant under $u$ displacements.
(iii) It is established that  there is 
exactly one  algebraic pure quasifree state which is invariant under $u$-displacements (of a fixed 
Bondi frame) and has positive self-adjoint generator 
of $u$-displacements. It coincides with the GNS-invariant state $\lambda$.
(iv) Finally it is showed that in the folium of a pure 
 $u$-displacement invariant state $\omega$ (like $\lambda$ but not necessarily quasifree) 
 on $\cW(\scri)$, $\omega$ is the only  
 state invariant under $u$-displacement.\\
(b) It is proved that the theory can formulated for spacetimes asymptotically flat at null infinity
which also admit  future  time completion $i^+$ (and fulfills other requirements 
related with global hyperbolicity). In this case a $*$-isomorphism $\imath$ exists - with a natural geometric meaning -
which identifies the (Weyl) algebra of observables of a linear field propagating in the bulk 
spacetime with a sub algebra of $\cW(\scri)$. Using $\imath$ a preferred state on the field algebra in the bulk spacetime
is induced by the $BMS$-invariant state $\lambda$ on $\cW(\scri)$.}

\normalsize

\section{Introduction.} 
\ssa{Summary the relevant results established in \cite{DMP} and some extensions}
Throughout $\bR^+:= [0,+\infty)$, $\bN:= \{0,1,2,\ldots\}$. For  smooth manifolds $M,N$, $C^\infty(M;N)$
(omitting $N$ whenever $N=\bR$) is the space of smooth functions $f: M\to N$.
$C^\infty_0(M;N)\subset C^\infty(M;N)$ is the subspace of compactly supported functions. 
If $\chi : M\to N$ is a diffeomorphism, $\chi^*$ is the natural extension to tensor bundles 
(counter-, co-variant and mixed) from $M$ to $N$ (Appendix C in \cite{Wald}).
A spacetime is a four-dimensional semi-Riemannian (smooth if no specification is supplied) 
manifold $(M,g)$, whose metric has signature $-+++$,
and it is assumed to be oriented and time oriented.
We adopt definitions of causal structures of Chap. 8 in \cite{Wald}.
  If $S\subset M\cap \tM$, $(M,g)$ and $(\tM,\tg)$ being spacetimes, $J^\pm(S;M)$ ($I^\pm(S;M)$) and $J^\pm(S;\tM)$
 ($I^\pm(S;\tM)$) indicate the causal (chronological) sets associated to $S$ and respectively referred to the spacetime 
 $M$ or $\tM$. 
  In \cite{DMP} we have considered a Weyl algebra constructed 
on the future null boundary of a {\em vacuum spacetime
asymptotically flat at  future null infinity} $(M,g)$.  
Following \cite{Wald},
a smooth spacetime $(M,g)$ is called {\bf asymptotically flat vacuum spacetime at future null infinity} if there is 
a second smooth spacetime  $(\tilde{M},\tilde{g})$ such that $M$ turns out to be an open
submanifold of $\tilde{M}$ with boundary $\scri \subset \tM$.  $\scri$ is an embedded 
submanifold of $\tM$ satisfying $\scri \cap J^-(M; \tilde{M}) = \emptyset$.
$(\tM,\tg)$ is required to be strongly causal in a neighborhood of $\scri$ and 
it must hold $\tilde{g}\spa\rest_M= \Omega^2 \spa\rest_M g\spa\rest_M$ where $\Omega \in C^\infty(\tM)$
is strictly positive on $M$. On $\scri$ one must have $\Omega =0$ and $d\Omega \neq 0$. 
Moreover, defining $n^a := \tg^{ab} \partial_b \Omega$, 
there must be a smooth function, $\omega$, defined in $\tM$ with $\omega >0$ on $M\cup \scri$, such that 
$\tilde{\nabla}_a (\omega^4 n^a)=0$ on $\Im$ and the integral lines of $\omega^{-1} n$ are complete on $\scri$.
Finally the topology of  $\scri$ must be that of $\bS^2\times \bR$. $\scri$ is called {\em future infinity} of $M$.\\
Hence $\scri$ is a 
$3$-dimensional submanifold of $\tilde{M}$ which is the union of integral lines of the null field 
$n^\mu:= \tilde{g}^{\mu\nu}\nabla_\nu \Omega$ (with $n\neq 0$  on $\scri$)
 and it is equipped with a degenerate metric $\tilde{h}$ induced by $\tilde{g}$.
The given definition is obtained by relaxing the original Ashtekar's definition \cite{AO}
 of {\bf vacuum spacetime asymptotically flat  at null and spatial infinity}, where the null infinity includes the 
 {\em past} null infinity $\scrip$ defined analogously to 
$\scri$. The spatial infinity is given by a special point in $\tM$ indicated by $i^0$ 
 (see Appendix \ref{infinities} and  Chapter 11 in \cite{Wald} for a general discussion). The results presented in
 \cite{DMP}
 does not require such a stronger definition.
For  brevity, from now on 
{\bf asymptotically flat spacetime} means  {\em vacuum spacetime asymptotically flat at future null infinity}.  
\remark Concerning this work,  vacuum Einstein 
equations need not to be valid everywhere on $M$, 
it is enough if they hold in a neighborhood of $\scri$ or, more weakly, 
``approaching'' $\scri$ as discussed on p.278 of \cite{Wald}.

As far as the only structure on $\scri$ is concerned,  
changes of the unphysical spacetime $(\tilde{M}, \tilde{g})$ associated with a fixed asymptotically flat spacetime $(M,g)$, are
completely encompassed
by {\bf gauge transformations} $\Omega \to \omega \Omega$ valid in a neighborhood of $\scri$,  
with $\omega$ smooth and 
strictly positive. Under these gauge transformations the triple $(\scri,\tilde{h}, n)$ transforms as
\beq
\scri \to \scri \:,\:\:\:\:\: \tilde{h} \to \omega^2 \tilde{h} \:,\:\:\:\:\: n \to \omega^{-1} n \label{gauge}\:.
\eeq
If $C$ is the class of  the triples $(\scri,\tilde{h}, n)$ transforming as in (\ref{gauge})
for a fixed asymptotically flat spacetime,
there is no general physical principle to single out a preferred element in $C$.
On the other hand, $C$ is {\em universal} for all asymptotically flat spacetimes \cite{Wald}:
If $C_1$ and $C_2$ are the classes of 
triples associated respectively to $(M_1,g_2)$
and $(M_2,g_2)$, there is a diffeomorphism $\gamma: \scri_1 \to \scri_2$ such that for suitable $(\scri_1,\tilde{h}_1, n_1)\in C_1$
and $(\scri_2,\tilde{h}_2, n_2)\in C_2$, 
\beq
\gamma(\scri_1) = \scri_2 \:,\:\:\:\:\: \gamma^* \tilde{h}_1=\tilde{h}_2 \:,\:\:\:\:\:\gamma^* n_1=n_2 \nonumber\:.
\eeq
With an appropriate choice of $\omega$ such that $\tilde{\nabla}_a (\omega^4 n^a)=0$, 
explicitly required to exist in the very definition of 
asymptotically flat spacetime, and using the fact that in a neighborhood of $\scri$ vacuum Einstein's equations are fulfilled,
the tangent vector $n$ turns out to be that 
of {\em complete null geodesics} with respect to $\tg$ (see Sec. 11.1 in \cite{Wald}).  
$\omega$ is completely fixed by requiring that, in addition,
the non-degenerate metric on the transverse section of $\scri$
is, constantly along geodesics, the standard metric of $\bS^2$ in $\bR^3$. We indicate by $\omega_B$ and
$(\scri,\tilde{h}_B,n_B)$ that value of $\omega$ and the associated triple respectively. 
For $\omega = \omega_B$,
a {\bf Bondi frame} on $\scri$ is a global coordinate system   $(u,\z,\bz)$ on $\scri$, 
where $u\in \bR$ is an affine parameter of the complete 
null $\tilde{g}$-geodesics whose union is $\scri$ and $\z,\bz \in \bS^2 \equiv \bC \cup \{\infty\}$ are complex 
coordinates on the cross section of 
$\scri$: 
$\z= e^{i\phi}\cot(\theta/2)$ with $\theta, \phi$ usual spherical coordinates of $\bS^2$.  
With these choices, the metric on the transverse section of $\scri$ reads $2(1+\z\bz)^{-2}(d\z\otimes d\bz+ d \bz \otimes d\z)
= d\theta \otimes d\theta + \sin^2 \theta \: d\phi\otimes d\phi$.\\
 By definition $\chi: \scri \to \scri$ belongs to the {\bf BMS group}, $G_{BMS}$ \cite{Penrose, Penrose2, Geroch, AS},  if $\chi$ is a diffeomorphism and 
 $\chi^*\tilde{h}$ and $\chi^*n$ differ from  $\tilde{h}$ and $n$ at most by a rescaling (\ref{gauge}). These diffeomorphisms 
represent
 ``asymptotic isometries'' of $M$ in the precise sense discussed in \cite{Wald} and highlighted in Proposition 2.1 
 in \cite{DMP}. 
Henceforth, whenever it is not explicitly stated otherwise,
{\em we consider as admissible realizations
of the unphysical metric on $\scri$ only those metrics $\tilde{h}$ which are accessible 
from a metric with associate
 triple $(\scri,\tilde{h}_B,n_B)$, by means of 
a transformations in $G_{BMS}$ }.

In coordinates of a fixed Bondi frame $(u,\z,\bz)$, the group  $G_{BMS}$ is realized as semi-direct group product 
$SO(3,1)\sp\uparrow \sp \times C^\infty(\bS^2)$, where
 $(\Lambda, f) \in SO(3,1)\sp\uparrow \times C^\infty(\bS^2)$ acts as
\begin{eqnarray}
u &\to & u':= K_\Lambda(\z,\bz)(u + f(\z,\bz))\:,\label{u}\\
\z &\to & \z' :=\Lambda\z:= \frac{a_\Lambda\z + b_\Lambda}{c_\Lambda\z +d_\Lambda}\:, \:\:\:\:\:\:
\bz \: \to \: \bz' :=\Lambda\bz := \frac{\overline{a_\Lambda}\bz + \overline{b_\Lambda}}{\overline{c_\Lambda}\bz +\overline{d_\Lambda}}\:.
\label{z}
\end{eqnarray}
$K_\Lambda$ is the smooth function on $\bS^2$
 \begin{eqnarray}
 K_\Lambda(\z,\bz) :=  \frac{(1+\z\bz)}{(a_\Lambda\z + b_\Lambda)(\overline{a_\Lambda}\bz + \overline{b_\Lambda}) +(c_\Lambda\z +d_\Lambda)(
 \overline{c_\Lambda}\bz +\overline{d_\Lambda})}
\label{K}\:\: \: \mbox{and}\:\:\:\:
 \left[
\begin{array}{cc}
  a_\Lambda & b_\Lambda\\
  c_\Lambda & d_\Lambda 
\end{array}
\right] = \Pi^{-1}(\Lambda)\:.
\end{eqnarray}
Above $\Pi$ is the well-known surjective covering homomorphism $SL(2,\bC) \to SO(3,1)\sp\uparrow$ (see \cite{DMP} for further
details). Two Bondi frames are connected each other through the transformations 
(\ref{u}),(\ref{z}) with $\Lambda \in SU(2)$. 
Conversely, any coordinate frame $(u',\z',\bz')$ on $\scri$ connected to a Bondi frame
 by means of an arbitrary
BMS transformation (\ref{u}),(\ref{z}) is {\em physically equivalent} to the latter 
from the point of view of General Relativity, but it is not necessarily a Bondi frame in turn.
A global reference frame $(u',\z',\bz')$ on $\scri$
related with a Bondi frame $(u,\z,\bz)$ by means of a  BMS transformation (\ref{u})-(\ref{z}) will be called
{\bf admissible frame}.
\remark The notion of Bondi frame is useful but {\em conventional}.  Any physical object must be invariant under 
the whole BMS group, i.e. under asymptotic symmetries  of $M$, and not only 
under the subgroup of $G_{BMS}$ connecting Bondi frames.

As in \cite{DMP}, let us consider asymptotically flat spacetimes $(M,g)$  satisfying the requirement that
there is an open set $\tilde{V}\subset \tM$  with 
$\overline{M\cap J^-(\scri;\tM)}\subset \tilde{V}$ (the closure being referred to $\tM$) such that $(\tilde{V}, \tg)$ 
 is globally hyperbolic. Under these hypotheses also $M_{\tilde{V}}:= \tilde{V}\cap M$ is globally hyperbolic.
The region in the future of a spacelike Cauchy surface of Schwarzschild and Minkowski spacetimes fulfill the requirement. (Vacuum spacetimes satisfying the requirement above which are 
also asymptotically flat at future and {\em past null, and spatial infinity}, 
 are called  {\em strongly asymptotically predictable} in the sense of Wald \cite{Wald}.
Minkowski spacetime is strongly asymptotically predictable.)  
If $\phi$ is smooth 
with compactly supported Cauchy data and
solves the  massless {\bf conformally-coupled Klein-Gordon equations} in $M_{\tilde{V}}$
 \beq P\phi=0\:, \quad \mbox{where $P:= -g^{\mu\nu}\nabla_\mu\nabla_\nu + \frac{1}{6}R$}\:, \label{KG}\eeq
  the limit $\psi$ of $(\omega_B\Omega)^{-1} \phi$ toward $\scri$ 
is smooth (Proposition 2.3 in \cite{DMP}).
 The action of asymptotic isometries on $\phi$ in the bulk corresponds to an action of $G_{BMS}$ on $\psi$  
 (Proposition 2.4 in \cite{DMP}) given by 
\beq
\left(A_{(\Lambda,f)}\psi\right)(u',\z',\bz') := K_\Lambda(\z,\bz)^{-1} \psi(u,\z,\bz)\label{alpha}
\eeq
 in a fixed 
 Bondi frame.
All that may suggest to think the rescaled boundary values $\psi$ as wavefunctions on $\scri$ and define a QFT 
based on a suitable symplectic space containing these wavefunctions where the BMS group, acting as in (\ref{alpha}),
is the symmetry group of the theory.
In fact, in \cite{DMP} we introduced a simple notion of QFT on $\scri$ based on a certain Weyl 
algebra of observables associated
with a symplectic space $(\cS(\scri), \sigma)$ with
 \beq
\cS(\scri) := \left\{ \psi \in C^\infty(\scri)\: \:\left|\:\: \sup_{(\z,\bz \in \bS^2)}|u|^{-k}|\partial^k_u \partial^m_\z
\partial^n_{\bz} \psi| \to 0\:, \mbox{as $|u|\to +\infty$}\:,  \forall k,m,n \in \bN \right.\right\} \:. \label{cSscri}
\eeq
 Here we {\em enlarge} $\cS(\scri)$ (the reason is validity of theorem \ref{holographicproposition2}) 
 up to the space $\sS(\scri)\supset \cS(\scri)$,
\beq
\sS(\scri) := \left\{ \left.\psi \in C^\infty(\scri)\: \:\right|\:\: \psi\: \mbox{and}\:\:  \partial_u 
\psi\:\:\mbox{belong to}\:\: L^2(\bR\times \bS^2, du \wedge \epsilon_{\bS^2}(\z,\bz)) \right\}  \label{Sscri}
\eeq
$\epsilon_{\bS^2}$  is (see below) the standard volume form of the unit $2$-sphere 
$\bS^2= \bC\cup \{\infty\}$. Both spaces are invariant under the action (\ref{alpha}) of $BMS$ group,
so that the choice of a Bondi frame in the definitions is immaterial.
Let us focus on the non-degenerate symplectic form $\sigma$. If $\psi_1,\psi_2 \in \cS(\scri)$ or, more generally
$\psi_1,\psi_2 \in \sS(\scri)$,
\beq
\sigma(\psi_1,\psi_2) := \int_{\bR\times \bS^2} 
\left(\psi_2 \frac{\partial\psi_1}{\partial u}  - 
\psi_1 \frac{\partial\psi_2}{\partial u}\right) 
du \wedge \epsilon_{\bS^2}(\z,\bz)\:, \quad  \epsilon_{\bS^2} (\z,\bz):= 
\frac{2d\z \wedge d\bz}{i(1+\z\bz)^2}  \:,\label{sigma}
\eeq
The Weyl algebra $\cW(\scri)$ is that associated with the pair $(\sS(\scri),\sigma)$ (see appendix \ref{algebras}). 
The generators of that Weyl algebras are denoted by $W(\psi)$, $\psi \in \sS(\scri)$. By definition they do not vanish 
and satisfy {\em Weyl relations} (or $CCR$)
$$(W1)\quad\quad W(-\psi)= W(\psi)^*\:,\quad\quad\quad\quad (W2)\quad\quad W(\psi)W(\psi') =
 e^{i\sigma(\psi,\psi')/2} W(\psi+\psi') \:.$$
$\cW(\scri)$ is uniquely determined, up to (isometric $*$-algebra) isomorphisms by the requirement that it is a
 $C^*$ algebra generated by non null elements
$W(\psi)$ fulfilling (W1) and (W2) (see Appendix \ref{algebras}).
The formal interpretation of generators is $W(\psi) = e^{i\sigma(\psi,\Psi)}$, 
$\sigma(\psi,\Psi)$ denotes the usual {\em symplectically smeared
field operator} (see appendix \ref{algebras}).\\
 Naturalness of the symplectic space $(\sS(\scri),\sigma)$ and the associated Weyl algebra 
is consequence of the following three facts. (i) $\sigma$  {\em is invariant
under the action (\ref{alpha}) of $BMS$ group}  as proved in Theorem 2.1 in \cite{DMP}, 
the enlargement of $\cS(\scri)$ does not affect the proof. 
(ii) Under suitable hypotheses, the Weyl algebra of linear QFT in the bulk identify with a sub algebra of 
$W(\scri)$. Let us enter this point with more details.
Let $(M,g)$ be an asymptotically flat spacetime such  that
there is an open set $\tilde{V}\subset \tM$  with 
$\overline{M\cap J^-(\scri)}\subset \tilde{V}$ and $(\tilde{V}, \tg)$ is globally hyperbolic. Define  $\cS_P(\MV)$ to be the real 
linear space of real smooth  solutions $\phi$ in $\MV$ of Klein-Gordon equation (\ref{KG}), which have compact support 
on Cauchy surfaces in $\MV$, and define the Cauchy-surface invariant symplectic form
\beq
\sigma_{M_{\tilde{V}}}(\phi_1,\phi_2) := \int_\Sigma \left(\phi_1 \partial_{n_\Sigma} 
\phi_2 - \phi_2 \partial_{n_\Sigma} \phi_1\right) d\mu^{(g)}_\Sigma\:, \quad \mbox{for $\phi_1,\phi_2 \in \cS_P(\MV)$,}\label{sigmaM}
\eeq
$\Sigma \subset M_{\tilde{V}}$ being a smooth spacelike Cauchy surface with unit, future directed, normal vector $n_\Sigma$
and measure $\mu^{(g)}_\Sigma$ induced by $g$.  In this context $\cW_P(\MV)$ denotes the Weyl algebra 
of the quantum field $\phi$ in the bulk associated 
with the symplectic space $(\cS(\MV), \sigma_{\MV})$.  Weyl generators are denoted by $W_{\MV}(\phi)$, $\phi \in \cS(\MV)$.
Proposition 4.1 in \cite{DMP} reads\\

\proposizione \label{holographicproposition}
{\em Let $(M,g)$ be an asymptotically flat spacetime such  that
there is an open set $\tilde{V}\subset \tM$  with 
$\overline{M\cap J^-(\scri;\tM)}\subset \tilde{V}$ and $(\tilde{V}, \tg)$ is globally hyperbolic.
Assume that both conditions below hold true for the  projection map 
$\Gamma_{\MV}: \cS_P(\MV) \ni \phi \mapsto \left((\omega_B \Omega)^{-1}\phi\right)\rest_{\scri}$:\\ 
{\bf (a)} $\Gamma_{\MV}(\cS_P(\MV))\subset \cS(\scri)$,\\
{\bf (b)} symplectic forms are preserved by $\Gamma_{\MV}$, that is, for all $\phi_1,\phi_2 \in \cS(\MV)$, 
\beq
\sigma_{\MV}(\phi_1,\phi_2) = \sigma(\Gamma_{\MV}\phi_1,\Gamma_{\MV}\phi_2)\:, \label{sigmam}
\eeq 
Then $\cW_P(\MV)$ can be identified with a sub $C^*$-algebra of $\cW(\scri)$
by means of a $C^*$-algebra isomorphism $\imath$ uniquely determined by the requirement
\beq
\imath(W_{\MV}(\phi)) = W(\Gamma_{\MV} \phi)\:, \:\:\:\:\mbox{for all $\phi\in \cS_P(\MV)$} \label{lc}\:,
\eeq}

\noindent By Proposition 4.1 in \cite{DMP} the conditions (a) and (b) are fulfilled  at least when $(M, g)$ is
the region in the future of a flat spacelike Cauchy surface in the  four-dimensional Minkowski spacetime
and $M_{\tV}=M$. In that case $\cS(\scri)$ (and thus $\sS(\scri)$) includes 
the limit $\psi$ to $\scri$ of the rescaled solutions $(\omega_B \Omega)^{-1}\phi$ of (\ref{KG})
in $M_{\tilde{V}}$ and $\sigma(\psi_1,\psi_2)$ coincides with the limit to $\scri$ of the bulk 
symplectic form. 
However, it is worth noticing that $(\sS(\scri),\sigma)$  does not depend on the particular  spacetime $M$ whose $\scri$ 
is the future causal boundary.

A preferred quasifree pure $BMS$-invariant state $\lambda$ on $\cW(\scri)$ has been introduced in \cite{DMP}.
The  extent is not affected by the enlargement of $\cS(\scri)$ to $\sS(\scri)$.
Fix a Bondi frame $(u,\z,\bz)$. For $\psi \in \sS(\scri)$ define its
{\bf positive-frequency part} $\psi_+$ (with respect to $u$) as follows:
\beq
 \psi_+(u,\z,\bz)  := \spa \int_{\bR}\spa e^{-iE u}\frac{\widetilde{\psi_+}(E,\z,\bz)}{\sqrt{4\pi E}} 
dE\:,\quad 
\frac{\widetilde{\psi_+}(E,\z,\bz)}{\sqrt{2E}} :=  \spa \frac{1}{\sqrt{2\pi}}
 \int_\bR\spa\spa e^{+iE u} {\psi}(u,\z,\bz)du\label{one}\:,
 \eeq  
with $\widetilde{\psi_+}(E,\z,\bz):=0$ for $E\not \in \bR^+$. 
With our enlargement of $\cS(\scri)$, the Fourier transforms in (\ref{one})
must be understood as the Fourier-Plancherel transforms (see Appendix \ref{Fourier}).
 From proposition \ref{last}, the right-hand side of (\ref{sigma}) can be computed 
also for positive frequency parties $\psi_{1+},\psi_{2+}$ when $\psi_1,\psi_2 \in \sS(\scri)$, provided the derivatives involved in (\ref{sigma})
be interpreted in distributional sense. A Hermitian scalar product arises in the 
complexified space of positive frequency parts:
\beq
\langle \psi_{1+},\psi_{2+} \rangle := -i \sigma(\overline{\psi_{1+}},\psi_{2+})\:. \label{is}
\eeq
Since $\sS(\scri)\supset \cS(\scri)$,
 Theorem 2.2 in \cite{DMP} implies that the Hilbert completion $\cH$ of the complexified space of positive frequency parts
 is isometrically isomorphic to $L^2(\bR^+\times \bS^2, dE\otimes \epsilon_{\bS^2})$ (no matter the enlargement 
 of $\cS(\scri)$).
In particular 
\beq
\al \psi_{1+},\psi_{2+}\cl = \int_{\bR^+\times \bS^2}\sp\sp \sp\sp\overline{\widetilde{\psi_{1+}}(E,\z,\bz)}\:
\widetilde{\psi_{2+}}(E,\z,\bz)\:dE\otimes\epsilon_{\bS^2}(\z,\bz)\:,\:\:\:
\mbox{for every pair $\psi_{1},\psi_{2} \in \sS(\scri)$.}\label{prodscalar2}
\eeq
Since $\sS(\scri)\supset \cS(\scri)$, Theorem 2.2 in \cite{DMP} implies also that the $\bR$-linear map $K : \sS(\scri)\ni \psi \mapsto \psi_+\in \cH$ 
 has dense range. Since,
by (\ref{one}) and (\ref{is}) one also has $\sigma(\psi_1,\psi_2) = -2 Im \langle K\psi_{1},K\psi_{2} \rangle$, we conclude 
(see proposition \ref{proposition2} in Appendix \ref{algebras}) that
there is a unique pure quasifree regular state $\lambda$ which satisfies, 
\beq
\lambda(W(\psi)) = e^{-\mu_\lambda(\psi,\psi)/2}\:, 
\quad \mbox{for all $\psi \in \sS(\scri)$ and where $\mu_\lambda(\psi_1,\psi_2):= Re \langle \psi_{1+},\psi_{2+} \rangle$} 
\label{lambda}
\eeq 
and the GNS triple $(\gH,\Pi,\Upsilon)$
is made of the Fock space $\gH$ with cyclic vector given by the vacuum $\Upsilon$ and one-particle space $\cH$. 
The representation $\Pi$ is completely determined by the identity, valid for every $\psi\in \sS(\scri)$,
 $\Pi\left(W(\psi)\right):= e^{i\overline{\Psi(\psi)}}$
where, following notation as in \cite{DMP} we write $\Psi(\psi)$ in place of 
$\sigma(\psi,\Psi)$ for the sake of simplicity.
\remark  {\em Througout this paper, the GNS triple of $\lambda$ and field operators, 
will be denoted omitting the index $_\lambda$}. 

Let us discuss on BMS invariance of the theory and the state $\lambda$. 
We recall the reader that a state $\omega$, on a $C^*$-algebra $\cA$, is  {\bf invariant} under a faithful 
$*$-automorphism representation $\beta$ of a group $G$, if 
$\omega(\beta_g(a)) = \omega(a)$ for every $g\in G$ and every $a\in \cA$.
Invariance of $\omega$ under $\beta$ implies that $\beta$ is {\em unitarily implementable} in the GNS representation 
$(\gH_\omega, \Pi_\omega, \Upsilon_\omega)$ of $\omega$ 
and there is a {\em unique unitary representation $U: G\ni g \mapsto U_{g}$ acting on $\gH_\omega$
  leaving fixed 
the cyclic vector} \cite{Araki}.
That is
\beq 
U_{g}\Pi_\omega(a)U_{g}^\dagger = \Pi_\omega\left(\beta_g(a)\right)
 \quad  \mbox{and} \quad U_{g} \Upsilon_\omega = \Upsilon_\omega\:, \quad \mbox{for all pairs $g\in G$, $a \in \cA$.}\label{U}
\eeq
The remaining unitary representations $\{V_g\}_{g\in \cG}$ of $\cG$ which implement the group on $\gH_\omega$ 
may transform $\Upsilon$
up to a phase $e^{ia_g}$ only. They therefore differ from $U$  for that phase at most\footnote{ 
$V_g \Pi_\omega(a)\Upsilon_\omega = e^{ia_g}\Pi_\omega(\beta_g(a))\Upsilon_\omega = e^{ia_g} U_g
\Pi_\omega(a)\Upsilon_\omega$.}.
When $G$ is a topological/Lie  group there is no guarantee, in general, for strong continuity of $U$
and thus for the existence
of self-adjoint generators, which, very often, have physical interest.\\
A group $G\ni g$ acting on a symplectic space $\sS$ by means of transformations $B_g$ 
preserving the symplectic form $\sigma$,
induces an analog $*$-automorphism representation $\beta$ on the Weyl algebra $\cW_{\sS,\sigma}$  
(see theorem 5.2.8 in \cite{BR2}). $\beta$ is
uniquely determined by $\beta_g(W(\psi)):=  W\left(B_{g^{-1}}\psi\right)$ for every $g\in G$ and $\psi\in \sS$. We call $\beta$
the {\bf representation canonically induced by} $G$. \\
Concerning $\cW(\scri)$,  $\sigma$ is  invariant under the action (\ref{alpha}) of $G_{BMS}$ and thus
 the representation $\alpha$, canonically induced by $G_{BMS}$ (\ref{alpha}) 
 on $\cW(\scri)$, is uniquely determined by the requirement 
\beq
\alpha_g(W(\psi)) = W\left(A_{g^{-1}} \psi\right)\:, \quad\mbox{for every $g\in G_{BMS}$ and $\psi\in \sS(\scri)$.} \label{auto}
\eeq
$\alpha$ turns out to be faithful.
With the extended definition of symplectic space we have the following theorem which embodies parts of
theorems 2.3 and 2.4 in \cite{DMP}.\\

\teorema \label{irrep} {\em The state $\lambda$ on $\cW(\scri)$ (\ref{lambda}) with GNS triple $(\gH, \Pi, \Upsilon)$
is invariant under the representation $\alpha$ of $G_{BMS}$ (\ref{alpha}), so that $\lambda$
is independent from the choice of the Bondi frame on $\scri$ used in (\ref{lambda}).
Furthermore the following holds.\\
{\bf (a)} The unique unitary representation $G_{BMS}\ni g \mapsto U_g$ representing $\alpha$ 
leaving fixed $\Upsilon$,  is the standard tensorialization of the representation $U^{(1)} = U\spa\rest_{\cH}$ 
on the one-particle
space $\cH$ defined  in the Bondi frame on $\scri$ used to define $\lambda$ by
\beq
  \left(U^{(1)}_{(\Lambda,f)}\varphi\right)(E,\z,\bz) = \frac{e^{iE K_{\Lambda}(\Lambda^{-1}(\z,\bz))f(\Lambda^{-1}(\z,\bz))}}{
  \sqrt{K_{\Lambda}(\Lambda^{-1}(\z,\bz))}}
 \varphi\left(E K_{\Lambda}\left(\Lambda^{-1}(\z,\bz)\right),\Lambda^{-1}(\z,\bz)\right) \:, \label{rep}
 \eeq
 for every $\varphi \in L^2(\bR^+\times \bS^2; dE \otimes \epsilon_{\bS^2})$ and $G_{BMS}\ni g \equiv (\Lambda,f)$.\\
{\bf (b)} $U$ is  strongly continuous when equipping $G_{BMS}$ with the nuclear topology (see \cite{DMP})}.\\

\noindent{\em Sketch of proof}. By direct inspection, referring to (\ref{one}), from (\ref{rep}) 
one sees that (i) $U^{(1)}$ is unitary and, if $\psi \in \sS(\scri)$,
$\psi^{(g)}_+(u,\z,\bz)  := \int_{\bR^+} 
\spa e^{-iE u}(U^{(1)}_g\widetilde{\psi_+})(E,\z,\bz)\frac{dE}{\sqrt{4\pi E}}$ is well defined and satisfies
(ii) $\psi^{(g)}_++ \overline{\psi^{(g)}_+} = A_g(\psi)$. Let $U_g$ be the tensorialization to the whole Fock space
of $U_g^{(1)}$ satisfying $U_g \Upsilon := \Upsilon$. Using $\Pi(W(\psi))= \exp\overline{i\sigma(\psi,\Psi)}$
(see proposition \ref{proposition2} in appendix \ref{algebras}), from (ii) arises 
$U_g \Pi(W(\psi)) U_g^\dagger = \Pi\left(W\left(A_{g^{-1}}\psi\right)\right)$. This proves (a) as well as 
 the invariance of $\lambda$ under $\alpha$ because $U_g \Upsilon := \Upsilon$ by constriction.
The proof of (b) is exactly that of Theorem 2.4 in \cite{DMP}. $\Box$\\

 \remark 
 It has been proved in Section 3 of \cite{DMP} that, adopting a suitable 
Wigner's-like representation analysis, (Theorem 3.2 in \cite{DMP})  the 
representation $U\spa \rest_{\cH}$ is that proper of a {\em massless} particle
with respect to the known {\em BMS notion of mass} \cite{Mc}. (The proof is completely independent
on the enlargement of $\cS(\scri)$ adopted here.)
 This is particularly relevant 
because this result suggests that, also in the absence of  Poincar\'e symmetry, the ``geometric
notion of mass'' which appears in Klein-Gordon equation could have a Schwinger - group theory  
interpretation, in relation to BMS group for asymptotically flat spacetimes. \\

\ssa{Contents of this paper} 
In this paper we primarily focus on one of the final  issues raised at the end of \cite{DMP}. {\em How is the BMS-invariant 
 state $\lambda$ unique?}
In fact, after some preparatory results given in section \ref{preparatory}, section \ref{mainsection} presents an answer to that question
based on some peculiarities of the state $\lambda$ which are examined in the following section.
In the practice, first we notice that  $\lambda$ enjoys positivity of the self-adjoint generator of $u$-translations
with respect to {\em every} admissible frame $(u,\z,\bz)$ on $\scri$. This fact may be interpreted 
as a remnant of spectral condition inherited from QFT in Minkowski spacetime. Moreover we find that,
every pure state on $\cW(\scri)$,  which is invariant under $u$-displacements with respect to a fixed 
admissible frame, satisfies cluster property with respect to these displacements.
Afterwards, in section \ref{mainsection} taking the cluster property into account,
  we show that the validity of positivity for the 
self-adjoint generator of $u$-translations
in {\em a fixed} admissible frame
individuates the BMS-invariant state $\lambda$ uniquely (without requiring BMS invariance).
As a second result, we show that, in the folium of a pure 
 $u$-displacement invariant state (like $\lambda$ but not necessarily quasifree) on $\cW(\scri)$, the state itself is the only  
$u$-displacement invariant state.
The proof of the first uniqueness result is essentially obtained by reducing to a uniqueness theorem due to Kay \cite{Kay}.

The second issue, considered in section \ref{mainsection2}, concerns the validity of proposition 
\ref{holographicproposition}
which assures that the Weyl algebra of a linear QFT in the bulk is isometrically mapped onto a sub algebra of $\cW(\scri)$.
We know that  proposition \ref{holographicproposition} holds for Minkowski spacetime. {\em Is that the only case?} The issue is important because the existence of the isometric $*$-homomorphism permits to 
induce a preferred state in the bulk by the symmetric state $\lambda$. We expect that the preferred state is invariant under any 
asymptotic symmetry (including proper symmetries) of the bulk by construction.
We prove in section \ref{mainsection2} which the isometric $*$-homomorphism of proposition \ref{holographicproposition}
exists whenever it is possible to complete $\scri$ by adding the asymptotic future point $i^+$ in the sense 
of Friedrich \cite{Friedrich}.

The last section contains some final comments and open questions.
 The appendices contain proofs of some propositions and recall general definitions and results used throughout.

\section{Some properties of $\lambda$, $\cW(\scri)$ and states on $\cW(\scri)$.} \label{preparatory}
\ssa{Positivity, $u$-displacement cluster property for Weyl-generator}
 There are two  interesting properties of $\lambda$ which were not mentioned in \cite{DMP}, these are stated 
in proposition \ref{positivity-clustering}.
 Some introductory notions are necessary.
For each admissible frame $\cF \equiv (u, \z,\bz)$ there is a one-parameter subgroup $\{\cT^{(\cF)}_t\}_{t\in \bR}$
of $G_{BMS}$ defining (active) $u$-displacements: $\cT^{(\cF)}:= \cT^{(\cF)}_t :(u,\z,\bz) \mapsto (u+t,\z,\bz)$. In turn, 
the restriction of $\alpha$ (\ref{auto}) to $\cT^{(\cF)}$ is a $*$-automorphism representation, 
$\alpha^{(\cF)}:= \{\alpha^{(\cF)}_t\}_{t\in \bR}$ of $\cT^{(\cF)}$.
An $\alpha^{(\cF)}$-invariant state $\omega$ on $\cW(\scri)$ is said to satisfy {\bf $\alpha^{(\cF)}$-cluster property 
for Weyl generators} if \beq
\lim_{t\to +\infty} \omega\left( W(\psi)\:\alpha^{(\cF)}_t\spa\left(W(\psi')\right)\right) = 
\omega\left(W(\psi)\right)\omega\left(W(\psi')
\right) \:, \quad\mbox{for all
$\psi,\psi'\in \sS(\scri)$.} \label{second}
 \eeq

\proposizione \label{positivity-clustering}
{\em The $G_{BMS}$-invariant state $\lambda$  on $\cW(\scri)$, defined in (\ref{lambda}), enjoys the following
properties with respect to the one-parameter group $\alpha^{(\cF)}$ of every admissible frame $\cF$.\\
{\bf (a)} The generator $H^{(\cF)}$ of the unitary group $\{e^{-itH^{(\cF)}}\}_{t\in \bR}$,  implementing $\alpha^{(\cF)}$
 leaving fixed the cyclic vector, is nonnegative.\\
{\bf (b)} $\lambda$ satisfies $\alpha^{(\cF)}$-cluster property  for Weyl generators (\ref{second}).}\\

\noindent {\em Proof}. It is sufficient to prove the thesis for a fixed Bondi frame $\cF$.
It generalizes to every other admissible frame $\cF'$ using the following facts: (i) $\lambda$ is $G_{BMS}$ invariant, 
 (ii) there is $g\in G_{BMS}$ such that
 $\alpha^{(\cF')}_t = \alpha_g\alpha^{(\cF)}_t \alpha_{g^{-1}}$ for every $t\in \bR$, 
 (iii) $\alpha$ is unitarily implementable leaving fixed the cyclic vector, 
 (iv) unitary equivalences preserve the spectrum of operators.\\
  (a)  Construct the state $\lambda$ referring to  the Bondi frame $\cF$. In the 
  one-particle space $\cH \equiv L^2(\bR^+\times \bS^2)$ of the GNS representation of $\lambda$, consider the self-adjoint
  operator $H$, such that 
  $(H \phi)(E,\z,\bz) := E\phi(E,\z,\bz)$
 defined in the domain of the square-integrable functions $\phi$ such that $\bR^+ \ni E\mapsto E\phi(E,\z,\bz)$
 is square integrable. $H$ has spectrum $\sigma(H):= [0,+\infty)$. By construction, if $H^\otimes$ denotes unique the standard
 tensorialization of $H$ extended to the Fock space $\gH$ with the constraint $H^{(\cF)}\Upsilon = 0$, 
 $H^{(\cF)}$ is non negative  by construction. Moreover 
  $e^{-itH^{(\cF)}}\Upsilon = \Upsilon$
  as well as, from (\ref{one}) 
 $$e^{-itH^{(\cF)}}\Pi\left(W(\psi)\right)e^{itH^{(\cF)}} = \Pi\left(W(A_{\cT^{(\cF)}_{-t}}\psi)\right) =
 \Pi\left(\alpha_{\cT_t^{(\cF)}}\left(W(\psi)\right)\right) = \Pi\left(\alpha^{(\cF)}_{t}\left(W(\psi)\right)\right) \:.$$
 We conclude that $e^{-itH^{(\cF)}}$ implements  $\alpha^{(\cF)}_t$ leaving fixed $\Upsilon$ and has nonnegative generator.
  \\ (b) Take $\psi,\psi'\in \sS(\scri)$. If
 $\psi'_t(u,\z,\bz) := \psi'(u+t,\z,\bz)$, 
using Weyl relations, (\ref{formal2}), invariance of $\lambda$  under $\alpha^{(\cF)}$ and (ii) in lemma \ref{lemma1},
  one has
\beq \lambda\left( W(\psi)\:\alpha^{(\cF)}_t\spa\left(W(\psi')\right)\right) = 
e^{-\langle K\psi,K\psi'_t \rangle}\lambda\left(W(\psi)\right)\lambda\left(W(\psi')
\right) \:.\label{quasi}\eeq
By (\ref{is}) and  Fubini-Tonelli theorem:
$\langle K\psi,K\psi'_t \rangle = \int_{\bR^+} dE \: e^{-itE} \int_{\bS^2} \overline{\widetilde{\psi_{+}}(E,\z,\bz)}\:
 \widetilde{\psi'_{+}}(E,\z,\bz)\: \epsilon_{\bS^2}(\z,\bz)$
where the internal integral defines a $L^1(\bR^+,dE)$ function of $E$.
 Riemann-Lebesgue lemma implies that $\langle K\psi,K\psi'_t \rangle $ vanishes as $t\to +\infty$, so that (\ref{second}) holds true
from (\ref{quasi}).
 $\Box$\\

 \remark 
  Consider QFT in Minkowski spacetime $\bM^4$ built up Minkowski vacuum $\Upsilon_{\bM^4}$
 and QFT with Weyl algebra $\cW(\bM^4)$ 
 on $\scri$ referred to $\lambda \equiv \Upsilon$. 
 If a Bondi frame $(u,\z,\bz)$ 
  on $\scri$ is associated with a Minkowski reference frame in the bulk, 
 $u$ displacements are in one-to-one correspondence with time translations respect  
 to the Minkowski frame. More precisely, by  Theorems 4.1 and  4.2 there is the unitary equivalence $\cU$
 which unitarily implements, in the respective GNS Hilbert spaces, 
 the $*$-isomorphisms $\imath: \cW(\bM^4) \to \cW(\scri)$ arising from proposition \ref{holographicproposition}
 (Proposition 4.1 in \cite{DMP}), 
 mapping $\Upsilon_{\bM^4}$
 into $\Upsilon$. Under the unitary equivalence $\cU$,  the self-adjoint generator of time displacements of the 
 Weyl algebra in the bulk is transformed to the self-adjoint generator of $u$-displacements for the
 Weyl algebra on $\cW(\scri)$. Hence the spectra of those operators are identical.
 Finally, as discussed in \cite{DMP}, changing Minkowski frame by means of a orthochronous Poincar\'e transformation 
 is equivalent to passing to another admissible frame (in general not a Bondi frame) 
 by means of a suitable transformation (\ref{u})-(\ref{z}).
 These changes preserve  the interplay of time displacements and $u$-displacements. We conclude that 
 positivity of $u$-generator for QFT on $\scri$ refereed to $\lambda$, valid for {\em every} admissible frame 
 on $\scri$, is nothing but
 the {\em spectral condition} of QFT in Minkowski spacetime referred to Minkowski vacuum for the free theory in $\bM^4$.
 In Minkowski QFT the spectral condition is a stability requirement: it guarantees that, under small (external) perturbations, the system 
 does not collapse to lower and lower energy states. In this way, we are lead to 
 consider positivity of $u$-displacement generator (with respect to {\em all} admissible frames on $\scri$) 
 as a natural candidate for replacing the spectral condition in QFT on $\scri$. We may assume it
 {\em also when $\scri$ is not thought as the null 
 boundary of Minkowski spacetime}.\\

\ssa{Asymptotic properties, extension of cluster property} 
The proof of proposition \ref{positivity-clustering}
yields, as a byproduct, a general property of $(\cW(\scri), \sigma)$, i.e. {\em asymptotic commutativity}.\\

\proposizione \label{asymptotic-commutativity} {\em  For every admissible frame $\cF$ the following facts are valid.}\\
{\em {\bf (a)} $\alpha^{(\cF)}$-{\bf asymptotic commutativity} holds:
\beq
\lim_{t\to +\infty} \left[\alpha_t^{(\cF)}\sp\left(a\right),\: b\right] =0\:, \quad \mbox{for all $a,b\in \cW(\scri)$.}
\label{three}\eeq
{\bf (b)} Let $\omega$ be a pure (not necessarily quasifree) state $\cW(\scri)$ with GNS 
representation  $(\gH_\omega, \Pi_\omega,\Upsilon_\omega)$ and assume that 
there exist   a unitary group $U^{(\cF)}$ implementing
$\alpha^{(\cF)}$ on $\gH_\omega$. Then $\omega$
satisfies 
$\alpha^{(\cF)}$-{\bf weak asymptotic commutativity}:
\beq
\mbox{w-}\lim_{t\to +\infty}  \left[U^{(\cF)}_tAU^{(\cF)\dagger}_t, B  \right] =0
\:, \quad\mbox{for all pairs $A\in \Pi_\omega(\cW(\scri))$,
$B\in \gB(\gH_\omega)$,}
\label{week}
\eeq 
where $\gB(\gH_\omega)$ is the space of 
bounded operators on $\gH_\omega$, 
 and  w-$\lim$ denotes  the weak operatorial topology limit.}\\
 
\noindent {\em Proof}. \label{proofs}
In the following $\cW_0$ is the $*$-algebra of finite linear combinations of all
$W(\psi)$, $\psi \in \sS(\scri)$.\\
(a)  First assume that $\cF$ is a Bondi frame and the coordinates of that Bondi frame 
to describe wavefunctions on $\scri$.
Using Weyl commutation relations one has, if $\psi'_t(u,\z,\bz):= \psi'(u+t,\z,\bz)$
$$||[\alpha_t^{(\cF)}\left(W(\psi')\right), W(\psi)]|| \leq 
|\sin \sigma(\psi'_t,\psi)|\:||W(\psi'_t+\psi)|| = |\sin(2 Im \langle K\psi,K\psi'_t\rangle)|\: ||W(\psi'_t+\psi)||\:,$$
where $K: \sS(\scri) \to \cH$ is that associated with the state $\lambda$. The left-hand side vanishes as $t\to +\infty$  
because
$||W(\phi)||=1$ for every $\phi \in \sS(\scri)$ and moreover, 
 we have seen in the proof of proposition 
\ref{positivity-clustering} that $\langle K\psi,K\psi'_t\rangle \to 0$ as $t\to +\infty$. 
If $\cF$ is not a Bondi frame, there is $g\in G_{BMS}$ such that, for every $t\in \bR$,
   $\alpha^{\cF}_t = \alpha_g \alpha^{\cF_0}_t \alpha_{g^{-1}}$ where $\cF_0$ is a Bondi frame. Using the fact that 
   $\alpha_g$ is a isometric $*$-automorphism which transforms a Weyl generators into a Weyl generator 
   and the result above, one gets $||[\alpha_t^{(\cF)}\left(W(\psi')\right), W(\psi)]||
   \to 0$ as $t\to +\infty$ again.
The result extends to $\cW_0$ by linearity. To conclude it is sufficient to extend the result to 
$\overline{\cW_0}= \cW(\scri)$. For every $\epsilon>0$ and fixed $a,b \in \cW(\scri)$ 
there are $a_\epsilon, b_\epsilon \in \cW_0$ with $||a-a_\epsilon||<\epsilon$, $||b-b_\epsilon||<\epsilon$.
Consider any sequence $t_n \to +\infty$. Since $\alpha^{(\cF)}$ is isometric 
 $0\leq \liminf_n |[\alpha_{t_n}^{(\cF)}\sp\left(a\right),\: b]| \leq \limsup_n |[\alpha_{t_n}^{(\cF)}\sp\left(a\right),\: b]|$ 
and\\
$\limsup_n |[\alpha_{t_n}^{(\cF)}\sp\left(a\right),\: b]| \leq 2||a-a_\epsilon||\: ||b_\epsilon|| +
2||a-a_\epsilon||\: ||b-b_\epsilon|| + 2||a||\: ||b-b_\epsilon|| 
+ \limsup_n |[\alpha_{t_n}^{(\cF)}\sp\left(a_\epsilon\right),\: b_\epsilon]|.$
The last term on the right-hand side converges to $0$ whereas the remaining terms are arbitrarily 
small. Therefore $|[\alpha_{t_n}^{(\cF)}\sp\left(a\right),\: b]| \to 0$ for any sequence $t_n \to +\infty$, i.e.
  (\ref{three}) is valid.\\
  (b) For the sake of simplicity, we indicate by $\Pi$ and $\gH$ respectively the GNS representation $\Pi_\omega$ and 
GNS Hilbert space $\gH_\omega$.
Since $\omega$ is pure, $\Pi(\cW(\scri))$ is irreducible. As a consequence, its commutant 
$\Pi(\cW(\scri))'$ contains only the elements $cI$ with $c\in \bC$. Thus the double commutant $(\Pi(\cW(\scri))')'$ 
coincides with $\gB(\gH)$.
Finally applying double commutant von Neumann's theorem, for that 
$\overline{\Pi(\cW(\scri))}^{\:s}= (\Pi(\cW(\scri))')'$,
 we conclude that $\overline{\Pi(\cW(\scri))}^{\:s}  = \gB(\gH)$,
$\overline{X}^{\:s}$ denoting the closure in the strong topology on $\gB(\gH)$ of any $X\subset \gB(\gH)$.
To go on, fix $a\in \cW(\scri)$, $B\in \gB(\gH)$ and 
take $\Psi_1,\Psi_2 \in \gH$. By $\overline{\Pi(\cW(\scri))}^{\:s}  = \gB(\gH)$, for each $\epsilon>0$ there is 
 $b_\epsilon \in \cW(\scri)$ with $||(B-\Pi(b_\epsilon))\Psi_1||<\epsilon$ and $||(B-\Pi(b_\epsilon))\Psi_2||<\epsilon$. With those
  choices, also exploiting  the fact that 
$||\Pi(\alpha_t^{(\cF)}\sp(a))|| \leq ||\alpha_t^{(\cF)}\sp(a)|| =||a||$, one has
$$\left|\left\langle \Psi_1, \left[\Pi\left(\alpha_t^{(\cF)}\sp\left(a\right)\right), B  \right] \Psi_2 \right\rangle\right|
\leq  \left|\left\langle \Psi_1, \Pi\left(\left[\alpha_t^{(\cF)}\sp\left(a\right), b_\epsilon  \right] \right)
\Psi_2 \right\rangle\right| + \epsilon ||a||\: (||\Psi_1||+ ||\Psi_2||)$$
Now, employing asymptotic commutativity and continuity of $\Pi$, one  concludes that the first term on the right-hand side
vanishes as $t\to +\infty$. Since $\epsilon>0$ is arbitrarily small, adapting the procedure, based on standard properties of 
$\limsup$ and $\liminf$, used in the proof of the item (a), one obtains that the limit of the left-hand side of the inequality above 
vanishes as $t\to +\infty$. $\Box$\\

\noindent To conclude this technical subsection,  we give a final proposition  which  
 extends $\alpha^{(\cF)}$-cluster property to the whole Weyl algebra establishing  also another related property. 
 If $\cF$ is a Bondi frame on $\scri$, we say that a state $\omega$ (not necessarily quasifree) on $\cW(\scri)$ satisfies 
$\alpha^{(\cF)}$-{\bf cluster property} (in the full-$\cW(\scri)$ version) if 
\beq \lim_{t\to +\infty}\omega(a\:\alpha_t^{(\cF)}(b)) = \omega(a)\omega(b)\:, \quad \mbox{for all $a,b \in \cW(\scri)$.}
\label{clusterfull}\eeq

\proposizione \label{main2}  {\em  Let $\cF$ be an admissible frame on $\scri$ and
 $\omega$ a pure (not necessarily quasifree) state on $\cW(\scri)$  
with GNS triple  $(\gH_\omega, \Pi_\omega,\Upsilon_\omega)$. If $\omega$
is  $\alpha^{(\cF)}$-invariant the following holds.\\
{\bf (a)} $\omega$ satisfies $\alpha^{(\cF)}$-cluster property.\\
  {\bf (b)}  If $A\in \Pi_\omega(\cW(\scri))$ and $U^{(\cF)}$ is a unitary group implementing $\alpha^{(\cF)}$ on $\gH_\omega$,
 one has
 \beq
 \mbox{w-}\lim_{t\to +\infty} U^{(\cF)}_t A U^{(\cF)\dagger}_t = \langle \Upsilon_\omega,A \Upsilon_\omega \rangle I\:.
 \eeq}
{\em  Proof}
(a) is an immediate consequence of (b) when writing the statement (a) in the GNS space $\cH_\omega$ using GNS theorem,
 with $A= \Pi_\omega(a)$ and $B=\Pi_\omega(b)$.
To  prove (a), take $B\in \Pi_\omega (\cW(\scri))$ and $\Phi \in \gH_\omega$. 
If $A_t = U^{(\cF)}_t AU^{(\cF)\dagger}_t$ and $P_0 = |\Upsilon_\omega\rangle \langle\Upsilon_\omega|$ we have
$$\langle \Phi, B A_t \Upsilon_\omega \rangle = \langle B^\dagger \Phi, [A_t, P_0]\Upsilon_\omega \rangle
+ \langle B^\dagger \Phi,  P_0A_t\Upsilon_\omega \rangle = \langle B^\dagger \Phi, [A_t, P_0]\Upsilon_\omega \rangle
+ \langle \Phi, B \Upsilon_\omega \rangle \langle \Upsilon_\omega, A_t \Upsilon_\omega \rangle\:.$$ 
The second term on the right-hand side is nothing but $\langle \Phi, B \Upsilon_\omega \rangle \langle \Upsilon_\omega, A
 \Upsilon_\omega\rangle$,
because $\omega$ is $\alpha^{(\cF)}$ invariant. Whereas the first term vanishes due to weak asymptotic commutativity. 
By asymptotic commutativity we also get
$\lim_{t\to +\infty}\langle \Phi, A_t B\Upsilon_\omega \rangle  = \lim_{t\to +\infty}\langle \Phi,B A_t 
\Upsilon_\omega \rangle =  \langle \Phi, B \Upsilon_\omega \rangle \langle \Upsilon_\omega, A \Upsilon_\omega \rangle$.
Since $\{B\Upsilon_\omega\}$ is dense in $\gH_\omega$, for every $\Psi \in \gH_\omega$ and $\epsilon>0$, there is 
$B_\epsilon\in \Pi_\omega (\cW(\scri))$ with $||B_\epsilon \Upsilon_\omega- \Psi|| < \epsilon$. Therefore, if 
$\langle A\rangle: = \langle \Upsilon_\omega, A\Upsilon_\omega\rangle$, it results that 
$|\langle \Phi, A_t \Psi \rangle - \langle A\rangle \langle \Phi, \Psi \rangle|$ is bounded by
$| \langle \Phi, A_t B_\epsilon \Upsilon_\omega\rangle  - \langle A\rangle \langle \Phi, B_\epsilon \Upsilon_\omega\rangle|
+ |\langle \Phi, A_t (\Psi- B_\epsilon \Upsilon_\omega)\rangle  - \langle A\rangle \langle \Phi, (\Psi - B_\epsilon \Upsilon_\omega)
\rangle|$.
The first term tends to $0$ as $t\to +\infty$, whereas the second is bounded by $\epsilon ||\Phi||\:|\:||A_t|| + |\langle A\rangle|\:|
= \epsilon ||\Phi||\:|\:||A|| + |\langle A\rangle|\:|$.
Finally, with the procedure based on standard properties of $\limsup$, $\liminf$ 
used in the proof of theorem \ref{main} below, one gets $\lim_{t\to +\infty}\langle \Phi, A_t \Psi\rangle  = 
  \langle \Upsilon_\omega, A \Upsilon_\omega \rangle\langle \Phi, \Psi \rangle $. 
$\Box$
 \section{The uniqueness theorem.} \label{mainsection}
 
\ssa{The uniqueness theorem} Making profitable use of cluster invariance, we are able to establish that $\lambda$ is the unique quasifree 
pure state on $\cW(\scri)$ such that (1) $\alpha^{(\cF)}$ invariant  for an, arbitrarily chosen, admissible frame $\cF$,
 and (2) the self-adjoint generator for the unitary implementation of $\alpha^{(\cF)}$ is non negative.
 No requirement about the full BMS invariance
 is necessary.  Moreover, dropping the quasifree hypotheses, we show that, in the folium\footnote{The {\em folium} 
of an algebraic state $\omega$ is the convex body of the states which are representable
by means of either vector or density matrices in the GNS Hilbert space of $\omega$.} of a pure 
 $\alpha^{(\cF)}$-invariant state on $\cW(\scri)$, $\omega$ is the only $\alpha^{(\cF)}$-invariant
state. Below ``$BMS$-invariant'' for a state means ``invariant under the $*$-automorphism representation $\alpha$  (\ref{auto})
of $G_{BMS}$''.\\

\teorema \label{main} {\em Consider an arbitrary admissible frame $\cF$ on $\scri$.  \\
{\bf (a)} The BMS-invariant state $\lambda$ 
defined in (\ref{lambda}) 
is the unique pure quasifree state on $\cW(\scri)$ satisfying both:

(i) it is invariant 
under $\alpha^{(\cF)}$,

(ii) the unitary group which implements $\alpha^{(\cF)}$ leaving fixed the cyclic GNS vector is strongly continuous with nonnegative
generator.\\
 {\bf (b)}  Let $\omega$ be a pure (not necessarily quasifree) state on $\cW(\scri)$ which is $BMS$-invariant 
 or, more weakly,  $\alpha^{(\cF)}$-invariant.
$\omega$ is the unique state on $\cW(\scri)$ satisfying both:

(i) it is invariant 
under $\alpha^{(\cF)}$, 

(ii) it belongs to the folium of $\omega$.}\\

\noindent{\bf Remarks}. {\bf (1)}
The condition (ii) in (a) is equivalent to the requirement  that
there is a strongly-continuous unitary 
group $\{e^{-itH^{(\cF)}}\}_{t\in \bR}$ implementing $\alpha^{(\cF)}$,
 such that $\inf \sigma(H^{(\cF)}) \geq \langle \Upsilon,
H^{(\cF)} \Upsilon \rangle$,  $\Upsilon $ being the cyclic GNS vector.\\
{\bf (2)} From a general result in the appendix \ref{algebras}, strong continuity 
for the unitary group implementing $\alpha^{(\cF)}$ leaving the cyclic vector unchanged for a state $\omega$,
 is equivalent to  
continuity at $0$ of $\bR\ni t \mapsto \omega\left( W(\psi)\:\alpha^{(\cF)}_t\spa\left(W(\psi')\right)\right)$
for all $\psi,\psi'\in \sS(\scri)$.\\
{\bf (3)} The BMS group
admits exactly one 4 dimensional Abelian normal subgroup -- it is
the group of translations $T$. If one uses Bondi-frame, then the
rigid $u$-displacement is just the action of a time-translation
in $T$. Therefore, the statemet (a) can be formulated
invariantly without reference to any Bondi or admissible frames: \\
{\em There is
exactly one algebraic pure quasi-free state which is invariant
under the action of any one time translation and the generator the
 time translation is non-negative; this state coincides
with the BMS invariant state $\lambda$.}\\
This result  is then
the most natural generalizations of those established by I. Segal
in Minkowski space \cite{Segal}: He also used a single time translation to
establish uniqueness of Minkowski vacuum and the unique vacuum was then shown to be
invariant under the Poincare group.  Of course the generalization holds only for zero rest
mass fields. \\

\noindent {\em Proof of theorem \ref{main}}. 
(a) Consider a state $\omega$ invariant under a one-parameter group of $*$-automorphisms
$\alpha^{(\cF)}$, supposing that $\cF$ is a Bondi frame, and let us indicate by $\{U_{t}^{(\cF)}\}_{t\in \bR}$ the unique 
unitary group which implements $\alpha^{(\cF)}$ leaving the GNS cyclic vector $\Upsilon_\omega$ fixed.
From now on we represent wavefunctions in coordinates $(u,\z,\bz)$ of $\cF$.
Since  $\omega$ is quasifree, one has 
$\Pi_\omega(W(\psi)) = e^{i\Psi_\omega(\psi)}$ and thus, in particular, for every $x\in \bR$,
 $U_t^{(\cF)}e^{i\Psi_\omega(x\psi)}U_{-t}^{(\cF)}
  \Upsilon_\omega =e^{i\Psi_\omega(x\psi_t)}\Upsilon_\omega$, where $\psi_t(u,\z,\bz) := \psi(u+t,\z,\bz)$.
 Using the fact that $U_{-t}^{(\cF)} \Upsilon_\omega = \Upsilon_\omega$  and applying Stone theorem, it results
 $U_{t}^{(\cF)}a^\dagger(K_\omega\psi) \Upsilon_\omega = a^\dagger(K_\omega\psi_t) \Upsilon_\omega$. In other words, 
 the one-particle space $\cH_\omega$ is invariant under $U_{t}^{(\cF)}$ and its restriction to $\cH_\omega$,
 $V^{(\cF)} := U^{(\cF)}\spa\rest_{\cH_\omega}$,
 is unitary as well. Tensorialization of $V^{(\cF)}$, assuming also invariance of $\Upsilon_\omega$,
 produces a unitary representation of $\alpha^{(\cF)}$ which leaves $\Upsilon_\omega$ fixed. Thus it 
 must coincide with $U^{(\cF)}$. As a consequence we can restrict our discussion to the one-particle space $\cH_\omega$.
 The fact that the $U^{(\cF)}$ is strongly continuous with positive self-adjoint generator implies that
 $V^{(\cF)}$ is strongly continuous with positive self-adjoint generator $H_\omega$. 
 Notice also that, 
 if $\psi_t(u,\z,\bz):= \psi_t(u+t,\z,\bz)$,
 by construction, $V_{t}^{(\cF)} K_\omega \psi = K_\omega \psi_{t}$ for every $t\in \bR$ and $\psi\in \sS(\scri)$.\\
  Now consider the triple $(K_\omega, \cH_\omega, V^{(\cF)})$ associated with $\omega$ (where $K_\omega : \sS(\scri) \to
 \cH_\omega$ is the function in lemma \ref{lemma1} and proposition \ref{proposition2})
 and the analog for $\lambda$, $(K, \cH, V^{(\cF)})$. We want to reduce to use the following remarkable result due to Kay \cite{Kay}.\\
 \lemma \label{lemma-kay}
 {\em  Let $\cW_{\sS,\sigma}$ be a Weyl algebra equipped with a one-parameter group of $*$-automorphisms $\beta = \{\beta_t\}_{t\in \bR}$
 canonically induced by a one-parameter group of transformations $B= \{B_t\}_{t\in \bR}$ of $\sS$ which preserve the symplectic form $\sigma$.
 Suppose that, for $k=1,2$, there are triples $(K_k, \cH_k, V_k)$ where:
  $\cH_k$ are complex Hilbert spaces,  $K_k: \sS\to \cH_k$ and $V_k = \{V_{k\:t}\}_{t\in \bR}$ are strongly continuous one-parameters groups of unitary operators
 on $\cH_k$. Suppose  that the following holds as well.\\
 (a)  $K_k$ are $\bR$-linear with
 dense range and  $\sigma(\psi,\psi')= -2 Im \langle K_k\psi , K_k\psi'\rangle_\cH$, with $\psi,\psi' \in \sS$.\\
 (b) $V_{k\:t} K_k \psi = K_k B_t \psi$ for every $t\in \bR$ and $\psi\in \sS$.\\ 
 (c) The self-adjoint  generators $H_k$ of $V_k$ have nonnegative spectrum.\\
 (d) $\overline{Ran H_k} = \cH_k$.\\
   With these hypotheses there is a unitary operator $U: \cH_1 \to \cH_2$
  with $UK_1= K_2$.}\\
 
 \noindent Notice that (2) of proposition \ref{proposition2} implies that, under the hypotheses of lemma \ref{lemma-kay},
 the pure quasifree states $\omega_1$ and $\omega_2$, respectively individuated by $(K_1,\cH_1)$ and $(K_2,\cH_2)$, must coincide.
Turning back to the proof of theorem \ref{main}, 
the triples $(K_\omega, \cH_\omega, V^{(\cF)})$ and $(K, \cH, V^{(\cF)})$ satisfy hypotheses (a) by lemma \ref{lemma1} 
 and (d) in proposition \ref{proposition2}. (b) and (c) hold true by construction/hypotheses for $\omega$ and
  by proposition \ref{positivity-clustering} for $\lambda$. 
 To conclude the proof of theorem \ref{main}
 it is now sufficient to establish the validity of (d), i.e. that $\overline{Ran H_\omega}= \cH_\omega$ and
 the analog for the generator $H$ of $V^{(\cF)}$.  
 Since  $H_\omega, H$ are self-adjoint, it is equivalent to prove that $Ker H_\omega = \{0\}$
 and $Ker H = \{0\}$.
 It is trivially true for the generator $H$ (see the proof of proposition \ref{positivity-clustering}).
 Let us prove that $Ker H_\omega = \{0\}$
 from cluster property, which is valid for $\omega$ due to (a) of proposition \ref{main2}. 
Dealing with as in (b) in the proof of proposition \ref{positivity-clustering} one 
 obtains
 \beq\omega\left( W(\psi)\:\alpha^{(\cF)}_t\spa\left(W(\psi')\right)\right) = e^{-\langle K_\omega\psi,K_\omega\psi'_t\rangle} 
 \omega\left(W(\psi)\right)\omega\left(W(\psi')
\right) \:.\label{bastarda}\eeq
Since $\omega\left(W(\psi)\right) = e^{-\mu_\omega(\psi,\psi)/2}\neq 0$ for every $\psi \in \sS(\scri)$, (\ref{bastarda})
together with cluster property,
  imply that $e^{-\langle K_\omega\psi,K_\omega\psi'_t\rangle} =1$ as $t\to +\infty$.
In other words
for every $\epsilon>0$ there is $T_\epsilon\in \bR$ with 
$$\left\langle K_\omega\psi, K_\omega\psi'_t\right\rangle  \in \bigcup_{n\in \bZ} B_\epsilon(2\pi i n)\:, \quad
\mbox{if $t> T_\epsilon$}\:,$$
where $B_\delta(\zeta):= \{z \in \bC \:|\: |z-\zeta|<\delta\}$.
However, the map $(T_\epsilon, +\infty) \ni t \mapsto \left\langle K_\omega\psi, K_\omega\psi'_t\right\rangle
=\left\langle K_\omega\psi,e^{-itH_\omega} K_\omega\psi'\right\rangle$
is continuous with connected domain and thus it must have connected range. Hence, if $\epsilon$ is small enough,
the range is contained in a single ball $B_\epsilon(2\pi i n_{\psi,\psi'})$. In turn, it implies
$$
\lim_{t\to +\infty}\frac{1}{2\pi i}\left\langle K_\omega\psi,e^{-itH_\omega} K_\omega\psi'\right\rangle =  n_{\psi,\psi'} \in \bZ 
\:, \quad\mbox{for all $\psi,\psi'\in \sS(\scri)$}\:.
$$
Linearity in $\psi$ implies that  $n_{\alpha \psi,\psi'}= \alpha n_{\psi,\psi'} \in \bZ$ for every $\alpha \in \bR$. Since 
$n_{\psi,\psi'} \in \bZ$,
it  is possible only if $n_{\psi,\psi'}=0$ for all $\psi,\psi'\in \sS(\scri)$ and hence:
$\left\langle K_\omega\psi,e^{-itH_\omega} K_\omega\psi'\right\rangle \to 0$ for all $\psi,\psi'\in \sS(\scri)$ if
$t\to +\infty$.
The result  extends to the whole space $\cH_\omega$. Indeed, if $\phi\in \cH_\omega$,
$$\left|\left\langle \phi,e^{-itH_\omega} K_\omega\psi'\right\rangle\right| \leq
\left|\left\langle K_\omega \psi,e^{-itH_\omega} K_\omega\psi'\right\rangle\right| + 
\left|\left\langle (\phi- K_\omega \psi),e^{-itH_\omega} K_\omega\psi'\right\rangle\right|\:,$$
now, using $||e^{-itH_\omega}||=1$,
$$0\leq \left|\left\langle \phi,e^{-itH_\omega} K_\omega\psi'\right\rangle\right| \leq  \left|\left\langle 
K_\omega \psi,e^{-itH_\omega} K_\omega\psi'\right\rangle\right| + ||\phi- K_\omega \psi||||K_\omega\psi'||\:.$$
As a consequence, for every sequence $\{t_n\}$ with $t_n\to +\infty$ as $n\to +\infty$
and for every $\psi \in \sS(\scri)$,
$$
0\leq \liminf_{n\to +\infty}|\left\langle K_\omega\psi,e^{-it_nH_\omega} K_\omega\psi'\right\rangle|\leq \limsup_{n\to +\infty}|\left\langle K_\omega\psi,e^{-it_nH_\omega} K_\omega\psi'\right\rangle| \leq ||\phi- K_\omega \psi||||K_\omega\psi'||\:,$$
As $\overline{Ran K_\omega}= \cH_\omega$, we can take $K_\omega \psi \to \phi$ in order to conclude  that, for every 
$\phi\in \cH_\omega$ and every $\psi'\in \sS(\scri)$:
$\left\langle \phi ,e^{-itH_\omega} K_\omega\psi'\right\rangle \to 0$ as $t\to +\infty$. Making use of the identity
$\left\langle \phi ,e^{-itH_\omega} K_\omega\psi'\right\rangle = 
\left\langle e^{itH_\omega} \phi , K_\omega\psi'\right\rangle$ and employing the same procedure, the result extends
to the right entry of the scalar product too. Summing up, cluster property yields
\beq
\lim_{t\to +\infty}\left\langle \phi,e^{-itH_\omega} \phi'\right\rangle = 0\:, \quad\mbox{for all $\phi,\phi'\in \cH_\omega$.}
\label{bastarda2}
\eeq
It is now obvious that, if there were $\phi_0\in Ker H_\omega \setminus \{0\}$ one would find
$\left\langle \phi_0,e^{-itH_\omega} \phi_0\right\rangle = \left\langle \phi_0, \phi_0\right\rangle \neq 0$ so that 
(\ref{bastarda2}) and cluster property, valid by proposition \ref{main2}, would be violated. Therefore $Ker H_\omega = \{0\}$. \\
Finally, we pass to consider the case where $\cF$ in the hypotheses is not a Bondi frame. Let  $\cF_0$ be a Bondi frame. 
There is $g\in G_{BMS}$ such that, for every $t\in \bR$,
   $\alpha^{(\cF)}_t = \alpha_g \alpha^{(\cF_0)}_t \alpha_{g^{-1}}$. The state $\omega'$ such that
   $\omega'(a):= \omega(\alpha_{g}(a))$ is invariant under $\alpha^{(\cF_0)}$ by construction.
   By direct inspection one sees
that the GNS triple of $\omega'$ is $(\gH_{\omega'}, \Pi_{\omega'}, \Upsilon_{\omega'}) =
(\gH_\omega, \Pi_\omega \circ \alpha_g, \Upsilon_\omega)$.
As a consequence, if $\{U_t\}_{t\in \bR}$ implements $\alpha^{(\cF)}$ for $\omega$ leaving $\Upsilon_\omega$ invariant, it also implements
$\alpha^{(\cF_0)}$ for $\omega'$ leaving fixed $\Upsilon_{\omega'}= \Upsilon_\omega$. Since, by hypotheses $\{U_t\}_{t\in \bR}$ is strongly 
continuous with positive generator and $\cF_0$ is a Bondi frame,
we can apply the result proved above for Bondi frames obtaining that $\omega' = \lambda$. That is $\omega\circ \alpha_g = \lambda$.
Since $\lambda$ is $BMS$ invariant, we have that $\omega= \lambda \circ \alpha_{g^{-1}}= \lambda$. \\
   (b).  Let $(\gH_\omega, \Pi_\omega, \Upsilon_\omega)$ be the GNS triple of a state $\omega$
 as in the hypotheses.
A generic element in the folium
of $\omega$ is a positive trace-class operator $\rho : \gH \to \gH$ with  $tr \rho =1$ and
has spectral decomposition $\rho = \sum_{i\in I} p_i |\Psi_i\rangle  \langle \Psi_i|$,
where $p_i \geq 0$ and $\sum_{i} p_i =1$. 
If $\rho \neq \lambda$ (i.e. $\rho \neq |\Upsilon \rangle \langle \Upsilon|$)  and 
$\rho$ is $\alpha^{(\cF)}$ invariant, 
the operator $P^\perp_0 \rho P^\perp_0/
tr(\rho P^\perp_0)$ 
($P_0^\perp$ denoting 
the orthogonal projector normal to $\Upsilon_\omega$)
is another well-defined  $\alpha^{(\cF)}$-invariant state in the folium of $\omega$.
Therefore, without loss of generality, we assume that each $\Psi_i$ in
$\rho = \sum_{i\in I} p_i |\Psi_i\rangle  \langle \Psi_i|$ satisfies 
 $\langle \Upsilon_\omega, \Psi_i\rangle=0$
 and we prove that every $p_i$ must vanish whenever $\rho$ is invariant under $\alpha^{(\cF)}$. 
Take $A = \Pi_\omega(a)$ with $a\in \cW(\scri)$ and let $A_t:= \Pi_\omega(\alpha_t^{(\cF)}(a))$. Since both $\omega$ and $\rho$
 are $\alpha^{(\cF)}$ invariant, one has:
 $$tr\left(\rho |A\Upsilon_\omega \rangle \langle A\Upsilon_\omega |\right) = tr\left(\rho |A_t\Upsilon_\omega \rangle 
\langle A_t\Upsilon_\omega |\right) 
 = \sum_{i\in I} p_i|\langle \Psi_i, A_t \Upsilon_\omega \rangle|^2 = 
 \lim_{t\to +\infty} \sum_{i\in I} p_i|\langle \Psi_i, A_t \Upsilon_\omega \rangle|^2$$
 $$= \lim_{t\to +\infty} \sum_{i\in I} p_i|\langle \Psi_i, P_0A_t \Upsilon_\omega \rangle + 
 \langle \Psi_i, [A_t, P_0] \Upsilon_\omega \rangle|^2 = \lim_{t\to +\infty} \sum_{i\in I} p_i|\langle \Psi_i, [A_t, P_0] \Upsilon_\omega \rangle|^2 =0$$
In the last step we used $\langle \Psi_i, [A_t, P_0] \Upsilon_\omega \rangle \to 0$ as $t\to +\infty$ due to weak asymptotic commutativity
of the state $\omega$. We have also interchanged the symbols of series and limit, using Lebesgue dominated convergence for the measure 
which counts the points of $I$. This is allowed by the $t$-uniform bound $|\:p_i|\langle \Psi_i, [A_t, P_0] \Upsilon_\omega \rangle|^2 \:|\leq p_i 2 ||a||$
and noticing that $\sum_{i\in I} 2 ||a|| p_i = 2||a||<+\infty$ by hypotheses. Since $\{A\Upsilon_\omega\}$ is dense in $\gH$ and
$0 = tr\left(\rho |A\Upsilon_\omega \rangle \langle A\Upsilon_\omega |\right) = \sum_{i\in I} p_i|\langle \Psi_i, A \Upsilon_\omega \rangle|^2$,
using a procedure based on Lebesgue's theorem again, one finds that $|\langle \Psi_i, \Psi_i \rangle|=0$  and thus 
$\Psi_i=0$ for every $i\in I$ as wanted.  This concludes the proof of (b). $\Box$\\

\section{Algebraic interplay bulk - $\scri$ in the presence of $i^+$ and induction of preferred states.} \label{mainsection2}
Proposition \ref{holographicproposition}
 assures that the Weyl algebra of a linear QFT in the bulk is isometrically mapped onto a sub algebra of $\cW(\scri)$,
 provided some hypotheses are fulfilled.
We know that the hypotheses of proposition \ref{holographicproposition} are fulfilled for Minkowski spacetime 
(more precisely the region in the future of a spacelike flat Cauchy surface therein). However the proof of the validity of these hypotheses for Minkowski spacetime, 
given in \cite{DMP}, exploited
 the fact that the {\em causal (Lichnerowicz') propagator} of the massless Klein-Gordon operator 
is strictly supported on the surface of the lightcone. It is known 
that, in general curved spacetimes, the support includes a "tail" supported inside the lightcone (this is equivalent to the generalized 
failure
of Huygens principle barring for ``plane-wave spacetimes'') \cite{gunther, friedlander}. In the following we show that, 
actually, the relevant hypotheses of proposition \ref{holographicproposition} and its thesis hold true  for 
the  class of spacetimes which are  flat at future null infinity (but not necessarily at past null and spatial infinity) and admit future 
time completion $i^+$ (once again Minkowski spacetime belongs to that class). The existence of such spacetimes in the class of vacuum 
solutions of Einstein equations was studied by Friedrich \cite{Friedrich} (actually his approach concerned spacetimes 
with {\em past} time completion $i^-$, but  re-adaptation to our case is immediate). Recasting the definition in \cite{Friedrich}
in a  language more useful for our goals, we have:\\

\definizione {\em A time-oriented four-dimensional smooth spacetime $(M,g)$ which solves vacuum Einstein equations 
is called {\bf asymptotically flat spacetime with future time infinity} $i^+$, if there is a smooth spacetime 
$(\tM,\tg)$ with a preferred point $i^+$, a diffeomorphism $\psi : M \to \psi(M) \subset \tM$  and a map $\Omega: \psi(M) \to [0,+\infty)$ so that
$\tg = \Omega^2 \psi^* g$ and the following facts hold. (We omit to write explicitly $\psi$ and $\psi^*$ in the following).\\
{\bf (1)} $J^-(i^+; \tilde{M})$ is closed and
$M = J^-(i^+)\setminus \partial J^-(i^+; \tilde{M})$.
(Thus $M= I^-(i^+; \tilde{M})$,  $i^+$ is in the future of
and time-like related with all the points of $M$ and $\scri \cap J^-(M; \tilde{M}) = \emptyset$.) 
Moreover $\partial M= \scri \cup \{i^+\}$ where
$\scri: = \partial J_-(i^+; \tilde{M}) \setminus \{i^+\}$ is the  {\bf future null infinity}. \\
{\bf (2)} $M$ is strongly causal.\\
{\bf (3)} $\Omega$ can be extended to a smooth function on $\tM$.\\
{\bf (4)} $\Omega\spa \rest_{\partial J_-(i^+; \tilde{M})} =0$, but $d\Omega(x) \neq 0$ for $x\in \scri$, and 
  $d\Omega(i^+) = 0$, but $\tilde{\nabla}_\mu \tilde{\nabla}_\nu \Omega(i^+) = -2 \tg_{\mu\nu}(i^+)$.\\
{\bf (5)}  If $n^\mu:= \tg^{\mu\nu} \tilde{\nabla}_\nu \Omega$, for a strictly positive smooth function $\omega$, defined in a neighborhood of $\scri$
 and satisfying $\tilde{\nabla}_\mu (\omega^4 n^\mu) =0$ on $\scri$,
the integral curves of $\omega^{-1}n$ are complete on $\scri$.}\\

\remark  
{\bf (1)} In \cite{Friedrich}, interchanging $i^+$ with $i^-$, the spacetimes defined above were called 
{\em vacuum spacetime with complete null cone at past infinity}.\\
{\bf (2)} As in the case of asymptotic flat spacetime at future null  infinity, 
the requirement that $(M,g)$ satisfies Einstein vacuum equations can be relaxed 
to the requirement that it does in a neighborhood of $\scri$ as far as one is interested in the 
geometric structure of $\scri$ only.\\
{\bf (3)} The conditions (4) and (5) were stated into a very different, but equivalent, form in \cite{Friedrich}.
In particular the last condition in (4) was required in terms of non degenerateness of the Hessian
$\tilde{\nabla}_\mu \tilde{\nabla}_\nu \Omega(i^+)$. It implies 
$\tilde{\nabla}_\mu \tilde{\nabla}_\nu \Omega(i^+) =  c g_{\mu\nu}(i^+)$ for some $c<0$ ($c>0$ in \cite{Friedrich} due to
the use of signature $+---$.) 
We fixed the value of the constant $c$, since global rescaling
of $\Omega$ are irrelevant. With our choices the null vector $\tilde{\nabla}^\mu \Omega$ is future directed along $\scri$.

 By (1), one has $J^-(M;\tM)\cap \scri = \emptyset$. Furthermore, dealing with as for the  analysis 
 performed in sec.11.1 in \cite{Wald} 
for asymptotically flat spacetimes,
the parts in conditions (4) and (5) referring to $\scri$, together with the fact that $(M,g)$ 
satisfies vacuum Einstein equations (in a neighborhood 
of $\scri$ at least) assure that $\scri$ is a smooth null $3$-surface (and embedded submanifold of $\tM$) 
made of the union of complete null
geodesics with respect to the metric $\omega^2 \tg$ (with $\omega$ as in requirement (5)) 
and that these geodesics are the integral curves  of $\omega^{-1}n$. Using the structure 
of the lightcone at $i^+$, where the metric $\tg$ is 
smooth, one sees that the topology of $\scri$ is $\bR\times \bS^2$.\\
 The gauge transformations (\ref{gauge}) and the BMS group have exactly the same meaning as in the case of  
 asymptotically flat spacetime (at future null infinity).
One can introduce the preferred gauge $\omega_B$,  Bondi frames and admissible frames once again.
Therefore BMS-invariant Weyl QFT based on $(\sS(\scri), \sigma)$ (with the preferred BMS invariant state $\lambda$)
can be recast as it stands for asymptotically flat spacetime with future time infinity too.\\
We come to the main result of this section. 
Let  $(M,g)$ be a {\em globally hyperbolic} asymptotically flat spacetime with future time infinity spacetime.
 Define  $\cS_P(M)$ to be the real 
linear space of real smooth  solutions $\phi$ in $M$ of Klein-Gordon equation (\ref{KG}) which have compact support 
on Cauchy surfaces in $M$. Define the Cauchy-surface invariant symplectic form 
\beq
\sigma_{M}(\phi_1,\phi_2) := \int_\Sigma \left(\phi_1 \partial_{n_\Sigma} 
\phi_2 - \phi_2 \partial_{n_\Sigma} \phi_1\right) d\mu^{(g)}_\Sigma\:, \quad \mbox{for $\phi_1,\phi_2 \in \cS_P(M)$,}\label{sigmaM'}
\eeq
$\Sigma \subset M$ being a smooth spacelike Cauchy surface with unit, future directed, normal vector $n_\Sigma$
and measure $\mu^{(g)}_\Sigma$ induced by $g$.  In this context $\cW_P(M)$ denotes the Weyl algebra 
of the quantum field $\phi$ in the bulk associated 
with the symplectic space $(\cS(M), \sigma_{M})$ with Weyl generators $W_{M}(\phi)$, $\phi \in \cS_P(M)$.

\remark $\cW_P(M)$ coincides with the algebra of {\em local observables} associated with the linear quantum field \cite{Wald2}. 
Localization is obtained by defining  Weyl generators $V(f):= W(Ef)$ (and field operators) smeared with smooth 
compactly-supported real functions $f: M\to \bR$. This is done by exploiting 
the {\em causal propagator} $E: C_0^\infty(M) \to \cS_P(M)$ (see the proof of the item (b)
in the thorem below). With this definition $[V(f),V(g)]=0$ if $supp f$ and $supp g$ are causally
separated (i.e. space-like related).\\

\teorema \label{holographicproposition2}
{\em Let $(M,g)$ be an asymptotically flat spacetime with future time infinity (where we remind the reader that, by definition,
 vacuum Einstein equations are supposed to be valid in a neighborhood of $\scri$).
Suppose that in the associated unphysical spacetime $(\tM,\tg)$ there is a 
open set $V \subset \tM$ with  $\overline{M} \subset V$ and  $(V,\tg)$ is globally hyperbolic. Then the  following facts hold.\\
{\bf (a)} $(M,g)$ is globally hyperbolic.\\
 {\bf (b)} The  projection map 
$\Gamma_{M}: \cS_P(M) \ni \phi \mapsto \left((\omega_B \Omega)^{-1}\phi\right)\rest_{\scri}$ is well-defined and satisfies
$\Gamma_{M}(\cS_P(M))\subset \sS(\scri)$.\\
{\bf (c)}
The symplectic forms are preserved by $\Gamma_{M}$, that is, for all $\phi_1,\phi_2 \in \cS_P(M)$, 
\beq
\sigma_{M}(\phi_1,\phi_2) = \sigma(\Gamma_{M}\phi_1,\Gamma_{M}\phi_2)\:.
\eeq 
{\bf (d)} $\cW_P(M)$ can be identified with a sub $C^*$-algebra of $\cW(\scri)$
by means of a $C^*$-algebra isomorphism $\imath$ uniquely determined by the requirement
\beq
\imath(W_{M}(\phi)) = W(\Gamma_{M} \phi)\:, \:\:\:\:\mbox{for all $\phi\in \cS_P(M)$} \label{lc2}\:.
\eeq
Thus, in particular, the BMS-invariant state $\lambda$ on $\cW(\scri)$ induces a quasifree state $\lambda_M$ on the field algebra of the bulk $\cW_P(M)$
by means of
\beq
\lambda_M(a) := \lambda(\imath(a))\:, \quad \mbox{for every $a \in \cW_P(M)$\:.} \label{state}
\eeq}

\noindent {\em Proof}. Without loss of generality we assume $V= \tM$.
We need a preliminary result given by the following lemma. 

\lemma \label{P1}{\em Consider a set $K\subset M$.
 In the hypotheses of the theorem  one has $J^-(K;M)= J^-(K;\tM)$
and $J^+(K;M)= J^+(K;\tM)\cap M$.}\\
{\em Proof}. To prove the identities, notice that every $\tM$-causal curve
completely contained in $M$ is a $M$-causal curve by construction. Therefore $J^-(K;M) \subset J^-(K;\tM)$
and $J^+(K;M) \subset J^+(K;\tM)\cap M$. To prove the former identity, suppose that there is  $s\in J^-(K;\tM)$
with $s\not\in J^-(K;M)$. $s$ must belong to an (at least continuous) $\tM$-causal 
past-directed curve $\gamma: [0,1] \to \tM$ 
 from $q\in K$ to $s$ which  includes points not contained in $M$. 
Since $\gamma(0)=q  \in  M$, the point $x = \gamma(t_x)$ such that 
$t_x= \sup\{t\in [0,1]\:|\: \gamma(u) \in M\:, \mbox{for $u\in [0,t]$}\}$
must belong to $\partial M = J^-(i^+; \tM)\setminus I^-(i^+; \tM)$. Notice also that, by construction
($q\in M = I^-(i^+;\tM)$) there must be 
a past-directed $\tM$-timelike line $\gamma'$ from $i^+$ to $q$.
If $x \in J^-(i^+; \tM)\setminus I^-(i^+; \tM)$  any (continuous) causal curves from $i^+$ to $x$ must be a portion of a 
smooth null a geodesic (Corollary to Theorem 8.1.2 in \cite{Wald}). In the considered case 
however, the continuous causal curve obtained by joining $\gamma'$ and $\gamma$ up to $x$ is a continuous causal curve and it is not
a portion of a null geodesic by construction. We conclude that $s$ cannot exist and $J^-(K;M)= J^-(K;\tM)$.
In the latter case, suppose that $s\in M$ satisfies  $s\in J^+(K;\tM)$,
but $s\not\in J^+(K;M)$. There must be at least one past-directed $\tM$-causal curve from $s$ to
$p\in K$ containing points in $\tM \setminus M$.   
In particular, as before, there is a past-directed causal curve $\gamma$ from $s\in I^-(i^+;\tM)$ to $x \in \partial J^-(i^+;\tM)$ 
and, in turn, there is a timelike 
past-directed curve from $i^+$ to $s$. By construction, the past-directed causal curve obtained by joining $\gamma'$ and $\gamma$ fails to be
a null geodesics,  so that $s$ cannot exists and hence $J^+(K;M)= J^+(K;\tM)\cap M$. $\Box$\\

\noindent {\em Proof of (a)}. Since $\tM$ is globally hyperbolic $J^+(p;\tM)\cap J^-(q;\tM)$ is compact 
(Theorem 8.3.10 \cite {Wald}). But one also has by lemma \ref{P1}
$J^+(p;\tM)\cap J^-(q;\tM)= J^+(p;\tM)\cap J^-(q;M) =
(J^+(p;\tM)\cap M)\cap J^-(q;M)) = J^+(p;M)\cap J^-(q;M)$ which, in turn, is compact as well. 
This is enough to establish that
$M$ is globally hyperbolic, it being strongly causal (see Remark at the end of Cap. 8 of \cite{Wald}).\\

\noindent {\em Proof of (b)}. Now we pass to consider causal (Lichnerowicz) propagators $E:= \Delta_--\Delta_+$ \cite{Dimok}, 
$\Delta_-$ and $\Delta_+$ being, respectively, the {\em advanced} and {\em retarded 
fundamental solutions} 
 associated with Klein-Gordon operator $P$ in a globally hyperbolic spacetime $N$. $\Delta_\pm : C_0^\infty(N) \to C^\infty(N)$ 
are uniquely defined by the requirements  that (i) they have the indicated domain and range, (ii) for every $f\in C_0^\infty(N)$, 
one has $P(\Delta_\pm f) =f$   with (iii) $\Delta_+f,\Delta_-f$ respectively supported in $J^+(supp f)$ and $J^-(supp f)$.
Now exploiting the fact that (see Appendix D of \cite{Wald}) in $M$ -- where $\Omega>0$ is smooth -- the following identity 
is fulfilled 
$$\Omega^{-3}(-g^{\mu\nu}\nabla_\mu\nabla_\nu + \frac{1}{6}R)\phi = (-\tg^{\mu\nu}\tilde{\nabla}_\mu\tilde{\nabla}_\nu + 
\frac{1}{6}\tilde{R})\Omega^{-1} \phi\:.$$ Furthermore, by lemma \ref{P1}, 
$J^-(supp f;M)= J^-(supp f;\tM)$ and $J^+(supp f;M)= J^+(supp f;\tM)\cap M$. In this way one easily gets that, if $f\in C_0^\infty(M)$ and  with obvious notation,
\beq \Omega(x)^{-1}\left(E f\right)(x) =   \tilde{E} (\Omega^{-3} f)(x)\:, \quad \mbox{for every $x\in M$.}\label{cp}\eeq
$\Box$\\

\noindent The proof of item (b) is obtained by collecting together the following three lemmata and taking into 
account the fact that  the standard measure of $\bS^2$, used in the definition of $\cS(\scri)$, is finite.\\

\lemma \label{P2}
 {\em $\Gamma_M \phi$ is well defined and is  a smooth function on $\scri$ for every $\phi \in \cS_P(M)$.}\\
{\em Proof}. Consider a smooth solution $\phi$ in $M$ of the equation $P\phi=0$ (\ref{KG}) with compactly supported Cauchy 
data, i.e. $\phi \in \cS_P(M)$. Then,
as $(M,g)$ is globally hyperbolic \cite{Wald2}, there is $C_0^\infty(M)$ with
$\phi = E f$. Since $\Omega^{-3}f\in C_0^\infty(M) \subset C_0^\infty(\tM)$,
we may also consider the solution  $\tilde{\phi} := \tilde{E}(\Omega^{-3}f)$ which is smooth and well defined in the 
whole globally hyperbolic spacetime 
$(\tM,\tg)$ and  on $\scri \cup \{i^+\}$ in particular. Due to (\ref{cp}), one has $\tilde{\phi}(x) = \Omega^{-1}(x) \phi(x)$ if $x\in M$. 
This proves $\Omega^{-1}\phi$ extends to a smooth function on $\tM$ and in particular to $\scri$. Since $\omega_B$ is smooth and 
strictly positive in a neighborhood of $\scri$, the analog holds considering $(\omega_B\Omega)^{-1}\phi$. $\Box$\\

\lemma \label{P3} {\em Referring to a Bondi frame $(u,\z,\bz)$ on $\scri$ and representing $supp(\Gamma_M \phi)$
in those coordinates, if $\phi \in \cS_P(M)$ there is $Q_\phi\in \bR$ with $supp(\Gamma_M \phi) 
\subset [Q_\phi,+\infty)\times \bS^2$.}\\
{\em Proof}.
Consider a Bondi frame $(u,\z,\bz)$ on $\scri$, {\em with $u$ future oriented}, and $\phi$ and  $f$ as above. 
$\omega\Gamma_M \phi = (\tilde{E}\Omega^{-3}f)\spa \rest_\scri = 
(\tilde{D}_-\Omega^{-3}f)\spa \rest_\scri - (\tilde{D}_+\Omega^{-3}f \spa)\rest_\scri$.
However  $(\tilde{D}_-\Omega^{-3}f \spa)\rest_\scri=0$ because $J^-(supp(\Omega^{-3}f); \tM) = J^-(supp(\Omega^{-3}f); M) \subset M = 
I^-(i^+; \tM)$.\\
Hence $supp(\Gamma_M \phi) = supp (\omega^{-1}\tilde{D}_-\Omega^{-3}f) \cap \scri = supp (\tilde{D}_-f) \cap \scri$ (in 
fact $\Omega^{-3}f$ and $f$
have equal support and $\omega>0$ on $\scri$).
Since $supp f$ is compact, there is a Cauchy surface $\Sigma$ for $\tM$
in the past of $supp f$ and in the past of $i^+$ and $supp(\Gamma_M \phi)= supp (\tilde{D}_-f) \cap \scri$ by consequence. 
Since $\Sigma$ and $\scri \cup \{i^+\} = \partial J^-(i^+; \tM)$ are closed,
 $S= \partial J^-(i^+;\tM) \cap \Sigma= \scri \cap \Sigma$ is such. The coordinate function $u: S \to \bR$ is smooth and in particular 
 continuous, so that it is bounded below on $S$ by some real $Q$. The same uniform bound holds for the $u$ coordinate 
of  the points in $supp (\Gamma_M \phi)= supp (\tilde{D}_-f) \cap \scri$, since $u$ is future oriented and those points are in the future 
of $S$. $\Box$\\
 
\lemma \label{P4}
{\em Consider a Bondi frame $(u,\z,\bz)$ on $\scri$. If $\phi \in \cS_P(M)$, for $p=0,1$, there is 
$u_0\in \bR$ sufficiently large and $C_p,M_p>0$,
such that, if $u>u_0$ and for every $(\z,\bz) \in \bS^2$
$$\left|\partial^p_u (\Gamma_M \phi)(u,\z,\bz)\right| \leq  \frac{M_p}{|C_pu-1|}\:.$$}
{\em Proof}. Since $\tM$ is globally hyperbolic, it is strongly causal. Consider a sufficiently small
open neighborhood $U$ of $i^+$ which is the image of exponential map centered at $i^+$ and consider $(U,\tg)$ as a spacetime. 
Strongly causality for $\tM$ implies that $I^-(i^+; U) = I^-(i^+;\tM)\cap U$. Therefore
$U \cap (\scri \cup \{i^+\}) = U \cap \partial I^-(i^+; \tM) = \partial^{(U)}(I^-(i^+;\tM)\cap U) = \partial^{(U)}I^-(i^+; U)$ 
(the topological boundary $\partial^{(U)} $ being referred to the topology of $U$). 
The structure of $\partial^{(U)} I^-(i^+; U)$ is 
well known (Theorem 8.1.12 in \cite{Wald}): it is made by the past-directed null geodesics through $i^+$.
Now consider Riemannian normal coordinates centered at $i^+$: $x \equiv (x^0,\bx)$ with $\bx := (x^1,x^2,x^3)$
 and defined in $U$ above. 
 From now on  $||\bx||:= \sqrt{(x^1)^2+(x^2)^2+(x^3)^2}$ and $|x| := \sqrt{(x^0)^2+(x^1)^2+(x^2)^2+(x^3)^2}$.
 In these coordinates
$\scri \cup \{i^+\}$ is the conical set $-x^0= ||\bx||$ and any geodesic through $i^+$ is 
a straight line $x^\mu(t) = c^\mu t$ for $t\in (-\epsilon, \epsilon)$ and $c^\mu\in \bR$ constants.
From now on we describe the portion of $\scri \cup \{i^+\}$ in $U$ by means of coordinates $(x^1,x^2,x^3)\in V$
where $V\subset \bR^3$ is open and bounded.
{\em We are explicitly assuming that $V$ includes $(0,0,0)$, corresponding to the tip of the cone $i^+$, where a conical singularity arises}
($\scri \cup \{i^+\}$ is not a submanifold of $\tM$ whereas $\scri$ is).
By direct inspection one sees that:\\
{\em if $f: \tM \to \bR$ is smooth, its restriction to  $\scri$
represented in function of $\bx$, $f\sp\rest_{\scri}(\bx):= f(-||\bx||,\bx)$, and  $\partial_{x^i}f\spa \rest_{\scri}(\bx)$ are smooth and bounded on $V\setminus \{(0,0,0)\}$ for $i=1,2,3$.
Boundedness generally fails for higher derivatives due to singularity of $||\bx||$ a the origin.}\\
Now consider an integral curve of $\tilde{\nabla}^\mu \Omega$, that is a solution of
$$\frac{dx^\alpha}{d\lambda} = \tg^{\alpha \beta} \partial_\beta \Omega(x(\lambda)) = -8 x^\alpha + O^\alpha_2(x)\:.$$
where we have used the conditions  $d\Omega(i^+)=0$ and $\partial_\alpha \partial_\beta \Omega(i^+)=
-2\tg_{\alpha\beta}(i^+)$ and 
the functions $O^\alpha_2$ satisfy
$O^\alpha_2(x)/|x|\to 0$ as $|x|\to 0$. As a consequence of standard theorems on dynamical systems, $x=(0,0,0,0)$
is a stable stationary point 
(the map $x \mapsto |x|^2$ being a Liapunov function for $i^+$)
and thus, for every $\epsilon>0$ there is $\delta>0$ such that 
 such that the integral lines satisfy, for all $\lambda >0$, $|x(\lambda)| <\epsilon$ if $|x(0)| <\delta$.
Multiplying both members of the differential equation for $x^\alpha$, summing over $\alpha$, and dividing for $|x(\lambda)|^2$ the result, one finally gets:
 $$\frac{d\ln|x(\lambda)|^2}{d\lambda} = -16 + \frac{O_2(x(\lambda))}{|x(\lambda)|}$$
 with $O_2(x)$ enjoying the same behaviour as $O_2^\alpha$ about $x=(0,0,0,0)$. Thus $|O_2(x(\lambda))|/|x(\lambda)|$ 
can be bounded from above by any
 arbitrarily small real $2\eta>0$, by taking the above-mentioned $\delta= 2\delta_\eta>0$ small enough. With this estimation one gets,
 if $|x(0)|<\delta_\eta$:
\beq
|x(\lambda)| \leq  2\delta_\eta e^{-(8-\eta) \lambda}\:. \label{est}
\eeq
$\tilde{\nabla}^\mu \Omega$ is a null vector tangent to $\scri$ (it can be seen by eq.(\ref{below}) by multiplying 
both sides by $\Omega$ and considering the limit as $\Omega=0$ i.e on $\scri$). Therefore integral lines with initial condition on 
$\scri$ belong to $\scri$ entirely. In this case (\ref{est}) produces, taking initial conditions with $||\bx(0)||<\delta_\eta$,
\beq
||\bx(\lambda)|| \leq  ||\bx(0)|| e^{-(8-\eta) \lambda} =  \delta_\eta e^{-(8-\eta) \lambda}\label{est2}\:.
\eeq
We come to the main issue. Let us consider a smooth function $\psi: \tM \to \bR$, in particular
the solution of Klein-Gordon equation in $(\tM, \tg)$ (which extend $\Omega^{-1} \phi$, $\phi$ being an associated
solution in $(M,g)$) considered. 
We want to evaluate the behaviour of $\omega_B^{-1} \psi\spa\rest_\scri$ in a neighborhood of $i^+$.
To this end we consider one of the above integral lines and the function (to be evaluated as $\lambda \to +\infty$)
$$\omega_B(\bx(\lambda))^{-1}\psi\spa\rest_{\scri}(\bx(\lambda))\:.$$
Barring re-arrangements in the cross section of $\scri$, $\omega_B$ is defined, along 
the considered integral lines,  by the equation (1.1.18)
of \cite{Wald}
$$
\frac{d\omega_B(\bx(\lambda))}{d\lambda} = -\frac{1}{2} 
\frac{\tg^{\mu\nu}\tilde{\nabla}_\mu \Omega \tilde{\nabla}_\nu \Omega}{\Omega}\rest_{\bx(\lambda)}\:.
$$
So that we have to study the behaviour at $\lambda \to +\infty$ of 
\beq \omega_B(\bx(\lambda)) = \omega_B(\bx(0)) e^{-\frac{1}{2}\int_{0}^{\lambda}
\frac{\tg^{\mu\nu}\tilde{\nabla}_\mu \Omega \tilde{\nabla}_\nu \Omega}{\Omega}\rest_{\bx(\lambda')} d\lambda'} \label{exp}\:.\eeq
The integrand is only apparently singular ($\Omega=0$ on $\scri$!) and it  must be evaluated using vacuum Einstein equations for $g$
$R_{\mu\nu}=0$, valid at least in a neighborhood of $\scri$, and employing
the conformal relation between Ricci tensor of $g$ and that of $\tg$:
\beq \Omega R_{\mu\nu} = \Omega\tilde{R}_{\mu\nu} + 2 \tilde{\nabla}_\mu \tilde{\nabla}_\nu \Omega +
\tg_{\mu\nu} \tg^{\alpha \beta}(\tilde{\nabla}_\alpha \tilde{\nabla}_\beta \Omega - 3 \Omega^{-1}
\tilde{\nabla}_\alpha \Omega\tilde{\nabla}_\beta \Omega) \label{below}\eeq
(see Eq. (11.1.16) of \cite{Wald}). For $\Omega=0$ (i.e. on $\scri$) one finds
\beq
 \frac{\tg^{\mu\nu}\tilde{\nabla}_\mu \Omega \tilde{\nabla}_\nu \Omega}{\Omega}\rest_{\scri} = 
 \tg^{\mu\nu}\tilde{\nabla}_\mu\tilde{\nabla}_\nu \Omega\rest_\scri\:.
 \eeq
The right-hand side tends to $-8$ as the argument approach $i^+$, because of the condition on $i^+$,
$\tilde{\nabla}_\mu \tilde{\nabla}_\nu \Omega(i^+) = -2g_{\mu\nu}(i^+)$. Using this result in (\ref{exp})
and (\ref{est})
we conclude that, for every $\epsilon>0$ we can choose a sufficiently small ball $B_{\delta_\epsilon}$  about $\bx= (0,0,0)$
containing all integral curves starting at $t=0$ inside this ball and such that, on these curves, for $\lambda \geq 0$,
$|8+\frac{\tg^{\mu\nu}\tilde{\nabla}_\mu \Omega \tilde{\nabla}_\nu \Omega}{\Omega}|\leq\epsilon$
so that
\begin{eqnarray}
\omega_B(\bx(0))e^{\lambda(4-\epsilon)}\leq  &\omega_B(\bx(\lambda))&\leq  \omega_B(\bx(0))e^{\lambda(4+\epsilon)}\label{1s}\:,\\
\quad (4-\epsilon)\omega_B(\bx(0))e^{\lambda(4-\epsilon)}\leq  &\frac{d\omega_B(\bx(\lambda))}{d\lambda}&\leq  (4+\epsilon)\omega_B(\bx(0))e^{\lambda(4+\epsilon)}
\label{2s}\:.
\end{eqnarray}
 Let  $\psi: \tM \to \bR$
the solution of Klein-Gordon equation in $(\tM, \tg)$ (which extend $\Omega^{-1} \phi$, $\phi$ being an associated
solution in $(M,g)$), if $|\psi\rest_\scri(\bx)|, |\partial_{x^i} \psi\spa\rest_\scri(\bx)| \leq M_\psi <+\infty$ in the 
considered ball $B_{\delta_\epsilon}$ about $\bx=(0,0,0)$
(and such a $M$ does exist as discussed at the beginning)
and $|\tg^{\alpha \beta} \partial_\beta \Omega(x(\lambda)) | \leq N <+\infty$ in $B_\delta$, (\ref{1s}) and (\ref{2s}) entail, for every $\lambda \geq 0$:
\begin{eqnarray}
\left|\omega_B(\bx(\lambda))^{-1} \psi\spa\rest_\scri(\bx(\lambda)) \right| &\leq&  
\frac{M_\psi}{\omega_B(\bx(0))}e^{-\lambda(4+\epsilon)}\label{1s'}\:,\\
\left|\frac{d}{d\lambda}\left(\omega_B(\bx(\lambda))^{-1} \psi\spa\rest_\scri(\bx(\lambda))\right) \right|
&\leq& \frac{M_\psi(4+\epsilon + 3N)}{\omega_B(\bx(0))}e^{-\lambda(4-3\epsilon)}\:.
\label{2s'}
\end{eqnarray}
To conclude we extend similar estimations to the case where the parameter of the integral curves is 
the coordinate $u$ of a Bondi frame. In this case the vector field to integrate is 
$\omega_B^{-1}(\bx(\lambda)) \tg^{\alpha \beta} \partial_\beta \Omega(x(\lambda))$ so that, along each curve,
$du/d\lambda = \omega_B(\bx(\lambda))$. As a consequence, integrating that equation making use of the estimation (\ref{1s}), one has 
in particular
\beq
 e^{-\lambda(4+\epsilon)} \leq \frac{\omega_B(\bx(0))}{(4+\epsilon)u - \omega_B(\bx(0))(3+\epsilon)} \label{us}\:.
\eeq
%\frac{\omega_B(\bx(0))}{(4-\epsilon)u - \omega_B(\bx(0))(3-\epsilon)} \leq e^{-\lambda(4-\epsilon)}\: \mbox{and}\:\:
As a consequence
\beq
 \left|\frac{\psi\spa\rest_\scri(\bx(u))}{\omega_B(\bx(u))} \right| &\leq&  
\frac{M_\psi}{(4+\epsilon)u - \omega_B(\bx_0)(3+\epsilon) }\label{1s''}\:.
\eeq 
Moreover
$$\left|\frac{d}{du}\left(\omega_B(\bx(u))^{-1} \psi\spa\rest_\scri(\bx(u))\right) \right|
= \frac{1}{\omega_B(\bx(u))} \left|\frac{d}{d\lambda}\left(\omega_B(\bx(\lambda))^{-1} \psi\spa\rest_\scri(\bx(\lambda))\right) \right|
\leq \frac{M_\psi(4+\epsilon + 3N)}{\omega_B(\bx(u))\omega_B(\bx_0)}e^{-\lambda(4-3\epsilon)}
$$
so that, by (\ref{1s}),
$$\left|\frac{d}{du}\left(\omega_B(\bx(u))^{-1} \psi\spa\rest_\scri(\bx(u))\right) \right| \leq 
\frac{M_\psi(4+\epsilon + 3N)}{\omega_B(\bx_0)^2}e^{-\lambda(8-4\epsilon)} \leq 
\frac{M_\psi(4+\epsilon + 3N)}{\omega_B(\bx_0)^2}e^{-\lambda(4+\epsilon)}\:.$$
Using (\ref{us}), one finally achieves
\beq
\left|\frac{d}{du}\frac{\psi\spa\rest_\scri(\bx(u))}{\omega_B(\bx(u))} \right| \leq 
\frac{M_\psi(4+\epsilon + 3N)}{\omega_B(\bx_0)[(4+\epsilon)u - \omega_B(\bx_0)(3+\epsilon)]} \label{f2}\:.
\eeq
Consider a ball $B_r$ centered in $\bx=(0,0,0)$ with radius $r<\delta_\epsilon$, so that 
all the estimation above are valid for the considered integral curves provided $\bx(\lambda=0)\in \partial B_r$. 
Referring to a Bondi frame $(u,\z,\bz)$, the coordinates $(\z,\bz)$ simply parametrize 
a class of the integral curves $\bx=\bx(u,\z,\bz)$. $\bx_0(\z,\bz)$ is the point, along the curve individuated by $(\z,\bz)$,
which belongs to $\partial B_r$. In global coordinates $(u,\z,\bz)$ on $\scri$, the  sphere $\partial B_r$ is represented 
as some compact surface, with equation $u= b(\z,\bz)$. For $u\geq b(\z,\bz)$, the integral line $\bx=\bx(u,\z,\bz)$
is completely contained in $B_r$ and thus (\ref{1s''}) and (\ref{f2}) are valid. Since $\omega_B$ is smooth 
and strictly positive, it attains its minimum $A>0$ and its maximum $B>0$ on the compact smooth manifold $B_r$.
As a consequence, inside $B_r$, i.e. for  $u> B$ and uniformly in $\z,\bz\in \bS^2$:
\beq
 \left|\frac{\psi\spa\rest_\scri(u,\z,\bz)}{\omega_B(u,\z,\bz)} \right| &\leq&  
\frac{M_\psi}{(4+\epsilon)u - B(3+\epsilon) }\label{x}\\
\left|\frac{\partial}{\partial u}\frac{\psi\spa\rest_\scri(u,\z,\bz)}{\omega_B(u,\z,\bz)} \right| &\leq& 
\frac{M_\psi(4+\epsilon + 3N)}{A[(4+\epsilon)u - B(3+\epsilon)]} \label{y}
\eeq 
These relations lead immediately to the thesis. $\Box$

\noindent Collecting together lemmata {\ref{P2}},  {\ref{P3}} and  {\ref{P4}}, one sees immediately that,
if $\phi \in \cS_P(M)$,
$\Gamma_M \phi$ is smooth and belongs to  $L^2(\bR\times \bS^2, du \wedge \epsilon_{\bS^2}(\z,\bz))$
together with its $u$-derivative because they have support included in a set $\{(u,\z,\bz) \in \bR \times \bS^2\:|\: u> Q\}$
for some $Q<+\infty$, decay sufficiently fast as $u\to +\infty$ and, finally, $\bS^2$ has finite (factor) measure.
In other words $\Gamma_M \phi \in \sS(\scri)$. This ends the proof of (b). $\Box$ \\

\noindent {\em Proof of (c)}. Consider $\phi_1, \phi_2 \in \cS_P(M)$ and a smooth spacelike 
Cauchy surface $\Sigma \subset M$.
If $K\subset S$ is compact and  includes Cauchy data of $\phi_1$ and $\Phi_2$, consider an open neighborhood $O\subset \Sigma$
with $O\supset K$. Then $V:= \overline{J^+(O; \tM) \cap M}$ (the closure being referred to $\tM$) 
includes the support of $\phi_1$ and $\Phi_2$ in the region between $\Sigma$ and $\scri \cup \{i^+\}$. One can arrange $V$ 
with local changes in order that 
the portion of $\partial V$ which does not intersect $\Sigma$ and $\scri \cup \{i^+\}$ is smooth. Notice that 
$\phi_i$ and $\phi_i/\Omega$ vanish smoothly on that portion of the boundary. 
If $S$ is a Cauchy surface for $\tM$ in the past of $O$, 
$V\subset D_+(S;\tM)\cap J^-(i^+;\tM)$ which is compact, therefore $\overline{V}$ is compact as well and has boundary 
smooth almost everywhere.
By direct inspection one finds that, if
$\sigma_M$ is defined as in (\ref{sigmam})
$$\sigma_M(\phi_1,\phi_2) = \int_\Sigma \left(\psi_1 \partial_{\tilde{n}_\Sigma} 
\psi_2 - \psi_2 \partial_{\tilde{n}_\Sigma} \psi_1\right) d\mu^{(\tilde{g})}_\Sigma$$
where, now, everything is referred to the unphysical metric $\tg = \Omega^2 g$ and
 $\psi_i := \Omega^{-1}\phi_i$. These fields are well defined solution of Klein-Gordon equation 
on $\tM$ and the right-hand side of the identity above coincides with the integral over $\Sigma$
of the $3$-form locally represented by
$$\chi_{\phi_1,\phi_2} := -\frac{1}{6}\sqrt{|\tilde{g}|}   \tg^{\gamma\alpha}  \left(\psi_1 \partial_\gamma 
\psi_2 - \psi_2 \partial_\gamma \psi_1\right)\epsilon_{\alpha \beta \mu \nu} dx^\beta \wedge dx^\mu \wedge dx^\nu\:,$$
($\epsilon_{\alpha \beta \mu \nu}$ is the sign of the permutation $\alpha \beta \mu \nu$ of $1234$ or 
$\epsilon_{\alpha \beta \mu \nu}=0$ if there are repeated numbers in $\alpha \beta \mu \nu$.)
We can use the divergence theorem for the form $\omega$ with respect to the region $\overline{V}$. As is well-known 
the fact that $\psi_i$ satisfies Klein-Gordon equation implies immediately that the divergence of $\omega$
vanishes. Since the boundary terms which are not evaluated on $\Sigma$ and $\scri\cup \{i^+\}$ do not give contribution,
the theorem of divergence reduces to the statement
\beq
\sigma_M(\phi_1,\phi_2)  = \int_{\scri} \chi_{\phi_1,\phi_2}\:.
\eeq
We have omitted $i^+$ since it has negligible measure (as is known an isolated conical singularity at the tip of a cone 
is too weak to create troubles with integration of smooth forms) and we assume that the orientation of $\scri$ is compatible with 
time orientation. It is known \cite{Wald} that $\scri$, $\Omega,u,\theta,\phi$ form a coordinate system in a full 
neighborhood of $\scri$ ($\theta,\phi$ are  standard coordinated on $\bS^2$) and that coordinate frame reduces to
a Bondi frame on $\scri$ for $\Omega=0$ with $\z= e^{i\phi}\cot(\theta/2)$. 
In these coordinates ((11.1.22) in \cite{Wald}, noticing that the metric therein is our $\omega_B^2 \tg$)
$$
\tg\spa\rest_\scri = \frac{-d\Omega \otimes du - du \otimes d\Omega + d\theta \otimes d\theta +\sin^2\theta\: d\phi \otimes d\phi}{\omega_B^2}\:.
$$
Since coordinates $u,\theta,\phi$
are adapted to $\scri$:
 $$\int_{\scri} \chi_{\phi_1,\phi_2} = -\int_{\bR \times \bS^2} \sqrt{|\tg\sp\rest_{\scri}|} \:\:\tg\sp\rest_{\scri}^{\Omega u} \left(\psi_1 \partial_u 
\psi_2 - \psi_2 \partial_u \psi_1\right) du \wedge d\theta \wedge d\phi\:.$$
Performing computations one has
$$\int_{\scri} \chi_{\phi_1,\phi_2} = -\int_{\bR \times \bS^2} \omega_B^{-2}\:\: \left(\psi_1 \partial_u 
\psi_2 - \psi_2 \partial_u \psi_1\right) du \wedge d\theta \wedge d\phi\:.$$
That is, since $\omega_B^{-2}\left(\psi_1 \partial_u 
\psi_2 - \psi_2 \partial_u \psi_1\right) =  \omega_B^{-1}\psi_1 \partial_u 
(\omega_B^{-1}\psi_2) - \omega_B^{-1}\psi_2 \partial_u (\omega_B^{-1}\psi_1)$, and passing to Bondi coordinates,
$$\sigma_M(\phi_1,\phi_2) = \int_{\bR \times \bS^2}  \left[\omega_B^{-1}\psi_1 \partial_u 
(\omega_B^{-1}\psi_2) - \omega_B^{-1}\psi_2 \partial_u 
(\omega_B^{-1}\psi_1)\right] du \wedge \frac{d\z \wedge d \bz}{i(1+ \z\bz)^2} = 
\sigma\left(\frac{\psi_1}{\omega_B},\frac{\psi_2}{\omega_B}\right)\:.$$
By the very definition of $\Gamma_M$, that is just the result we wanted to establish. $\Box$\\

\noindent {\em Proof of (d)}.
If $\cW(\scri)_M$ is the $C^*$-algebra of $\cW(\scri)$ generated by generators
$W(\Gamma_M\phi)$ for every $\phi \in \cS_P(M)$, preservation of symplectic forms by the linear map $\Gamma_M$
implies immediately  (theorem 5.2.8 in \cite{BR2}) that there is a unique (isometric) $*$-algebra isomorphism $\imath$
from $\cW_P(M)$ to $\cW(\scri)_M$ satisfying (\ref{lc2}). The statement concerning the induction of the state 
$\lambda_M$ is straightforward. In particular, 
the fact that the state is quasifree follows immediately from the expression (\ref{lambda}) for $\lambda$. It implies
that $\lambda_M$ is the quasifree state associated with the scalar product 
$\mu_M(\phi,\phi'):= \mu_\lambda(\Gamma_M \phi, \Gamma_M \phi')$. Preservation of symplectic forms
assures that $\mu_M$ fulfills (\ref{sm}) with respect to $\sigma_M$.
$\Box$\\
The proof of the theorem is concluded.
$\Box$\\

\remark To conclude, we notice that Minkowski spacetime $(\bM^4,\eta)$, more precisely the region of  $(\bM^4,\eta)$ 
in the future of a spacelike Cauchy surface representing the rest space of an arbitrary Minkowskian frame,
fulfills both  the definitions of 
asymptotic flat at future null infinity spacetime and asymptotically flat spacetime with future time infinity.
In particular in both cases $\scri$ is the same submanifold 
of $(\tM,\tg)$, the latter being  Einstein closed universe 
(see \cite{Wald,DMP}).  Since  Einstein closed universe is globally hyperbolic, 
theorem \ref{holographicproposition2} is valid in this case. However the thesis of the theorem is true anyway
because of the independent proof given in that case in (a) of Theorem 4.1 in \cite{DMP}. 
We also know by (b) of Theorem 4.1 in \cite{DMP} 
that, in the considered case, the 
state $\lambda_M$ induced by $\lambda$ is nothing but Minkowski vacuum.

\section{Discussion and open issues.} A crucial role in proving the uniqueness theorem was played by the fact that the $C^*$
algebra of observables is a Weyl algebra: this fact is essential in obtaining both cluster property for every state which is invariant 
under $u$-displacements and asymptotic commutativity, used in establishing the uniqueness theorem. The use   
of a Weyl algebra to describe quantum observables 
in standard QFT in a globally hyperbolic spacetime is appropriate as far as the theory deals with linear -- i.e. ``free'' -- fields.
This is because nonlinear field equations -- i.e. the presence of ``interaction'' -- do not preserve the standard 
symplectic form of field solutions if varying  Cauchy surface. However dealing with QFT on $\scri$, the extent is  
different since there is no time evolution -- one stays ``at the end of time'' when interactions of the bulk, if any,
have been switched off -- and a Weyl algebra may still be appropriate. It is especially if one try to use some
``$S$ matrix'' formalism (involving QFT on $\scrip$) in order to describe bulk phenomena in terms of features
of QFT on the boundary of the spacetime.
In this view, the state in the bulk spacetime induced from the unique
BMS-invariant state on $\scri$ whose existence is guaranteed in
         theorem \ref{holographicproposition2} should have, of course, the natural interpretation of an ``out
            vacuum state'', and thus theorem \ref{holographicproposition2} in particular establishes
            rigorously the existence of such an out vacuum state under the
            precise technical requirements about the nature of the
            asymptotics of the metric. However  if adopting this point of view, any outcoming $S$-matrix
 theory would enjoy a larger symmetry, based on the BMS group, rather than the usual Poincar\'e one.
 Finally, it should be emphasized that the unique vacuum state considered in theorem \ref{main}
looking at $\scri$ will not in general coincide with that picked,
by an identical construction, on $\scrip$, if the spacetime is asymptotically flat also at past null infinity, 
unless the spacetime is stationary. \\
 Concerning the last statement of theorem \ref{holographicproposition2}, we remark that 
the state $\lambda_M$ is universal: it does not depend on the particular asymptotic flat (with $i^+$) spacetime 
under consideration, but only on the fact that it is asymptotically flat.
An important issue 
 deserving further investigation is the validity of Hadamard property \cite{KW,Rad} for the state $\lambda_M$.
 In case this property is fulfilled, it make sense to implement a perturbative procedure 
 to study the back reaction on the metric using the stress-energy tensor operator \cite{stress} averaged 
 on $\lambda_M$. Failure of Hadamard property would imply dubious gravitational stability of the spacetime.
 A first scrutiny seems to shows that, at least near $\scri$, the singular support of the two-point function
 associated to $\omega_M$ is included in the set of couple of points connected by means of a null
 geodesic. This is a first clue for the validity of Hadamard behaviour.
 Another property of $\lambda_M$ which is, most probably fulfilled, is its symmetry with respect every
 proper isometry group of $M$ if any. This is because $\lambda$ is invariant under BMS group which includes
 (asymptotic) symmetries. A general open problem, which seems to be quite difficult for several 
 technical reasons, is the extension of the results presented here and in \cite{DMP} to the case of 
 a massive field. Al these issues will be investigated elsewhere. \\

\noindent{\bf Acknowledgments}. I am grateful to C. Dappiaggi and N.Pinamonti for several fruitful discussions
and suggestions. In particular, I would like to thank C. Dappiaggi for having pointed out
references \cite{Friedrich}.

\appendix

\section{Quasifree states on Weyl algebras} \label{algebras}
 A $C^*$-algebra $\cW_{(\sS,\sigma)}$ is called {\bf Weyl algebra} associated 
%%%CORREZIONE: frase aggiunta
with a (real) symplectic space $(\sS,\sigma)$ (the symplectic form $\sigma$ being nondegenerate) 
%%%
if it contains a class of non-vanishing elements $W(\psi)$ for all
 $\psi \in \sS$, called {\bf Weyl generators}, 
  satisfying {\bf Weyl relations}\footnote{Notice that in \cite{KW} a different convention for the sign of $\sigma$ in (W2) is employed.}:
$$(W1)\quad\quad W(-\psi)= W(\psi)^*\:,\quad\quad\quad\quad (W2)\quad\quad W(\psi)W(\psi') =
 e^{i\sigma(\psi,\psi')/2} W(\psi+\psi') \:;$$
 and $\cW_{(\sS,\sigma)}$ coincides with the closure of the $*$-algebra (finitely) generated
 by Weyl generators. $\cW_{(\sS,\sigma)}$ is {\em uniquely} determined by $(\sS,\sigma)$  (theorem 5.2.8 in \cite{BR2}): 
two different realizations admit a unique $*$ isomorphism which transform the former into the latter
preserving Weyl generators and the norm on $\cW_{(\sS,\sigma)}$ is unique since
$*$ isomorphisms of $C^*$ algebras are isometric. This results implies that every GNS representation of a Weyl algebra is always faithful
and isometric.
$\cW_{(\sS,\sigma)}$ can always be realized in terms of bounded operators on $\ell^2(\sS)$, viewing $\sS$ as a
Abelian group and defining the generators as
$(W(\psi)F)(\psi'):= e^{-i\sigma(\psi,\psi')/2}F(\psi+\psi')$ for every $F\in \ell^2(\sS)$.
In this realization (and thus in every realization) it turns out evident that generators $W(\psi)$ are {\em linearly independent}. 
As a consequence of (W1) and (W2), one gets: $W(0) = \bI$ (the unit element), 
$W(\psi)^*= W(\psi)^{-1}$, 
$||W(\psi)||=1$  and,  using non degenerateness of $\sigma$, $W(\psi)=W(\psi')$ 
iff $\psi=\psi'$. 
 Strong continuity of the unitary group implementing a $*$-automorphism representation
$\beta$ of a topological group $G \ni g \mapsto \beta_g$ for a $\beta$-invariant 
state  $\omega$ on a Weyl algebra 
$\cW(\sS,\sigma)$, 
is equivalent to $\lim_{g\to \bI} \omega(W(-\psi)\beta_g W(\psi)) = 1$ for all $\psi \in \sS$.
The proof follows immediately from  
$||\Pi_\omega\left(\beta_{g'}  W(\psi)\right)\Upsilon_\omega - \Pi_\omega\left(\beta_{g}  W(\psi)\right)\Upsilon_\omega||^2
= 2- \omega\left(W(-\psi)\beta_{g'^{-1}g}W(\psi)\right) -  
\omega\left(W(-\psi)\beta_{g^{-1}g'}W(\psi)\right)$
 and $\overline{\Pi_\omega(\cW(\sS,\sigma))\Upsilon_\omega} = \gH_\omega$.\\
 A state $\omega$ on $\cW_{(\sS,\sigma)}$, with GNS triple $(\gH_\omega, \Pi_\omega, \Upsilon_\omega)$, is called
{\bf regular} if the maps $\bR\ni t\mapsto \Pi_\omega(W(t\psi))$ are strongly continuous. 
Then, in accordance with Stone theorem, one can write 
$\Pi_\omega(W(\psi)) = e^{i\sigma(\psi,\Psi_\omega)}$,  
 $\sigma(\psi,\Psi_\omega)$ being the (self-adjoint) {\bf field operator symplectically-smeared} with $\psi$. 
In this way  field operators enters the theory in Weyl algebra scenario.
 Working formally, by Stone theorem (W2) 
 implies $\bR$-linearity and standard CCR:
$$
(L)\quad \sigma(a\psi + b \psi', \Psi_\omega) = a \sigma(\psi,\Psi_\omega) + b \sigma(\psi',\Psi_\omega)\:, \quad\:\: 
(CCR) \quad \mbox{$[$}\sigma(\psi,\Psi_\omega), \sigma(\psi',\Psi_\omega)\mbox{$]$} = -i\sigma(\psi,\psi')I\:,$$  
for $a,b\in \bR$ and $\psi,\psi' \in \sS$.
Actually (L) and (CCR) hold rigorously in an invariant dense set of analytic vectors
by Lemma 5.2.12 in \cite{BR2} (it holds if $\omega$ is quasifree by proposition 
\ref{proposition2}).\\
In the standard approach of QFT, based on bosonic real scalar field operators $\Psi$ a, {\em either vector or density matrix}, 
state is {\em quasifree} 
if the associated $n$-point functions (expectation values of a product of $n$ fields) satisfy
 (i) $\langle\sigma(\psi,\Psi) \rangle =0$ for all $\psi\in \sS$ and (ii)
 the  $n$-point functions  $\langle  \sigma(\psi_1,\Psi)\cdots \sigma(\psi_n,\Psi)\rangle$
are determined from the functions $\langle \sigma(\psi_i,\Psi)\sigma(\psi_j,\Psi) \rangle$, with
$i,j=1,2,\cdots, n$,
using standard Wick's expansion. 
A technically different but substantially equivalent definition, completely based on the Weyl algebra
 was presented in \cite{KW}. It relies on the following three observations.
(a)Working formally with (i) and (ii), one finds that it holds
$\langle  e^{i\sigma(\psi,\Psi)} \rangle =
 e^{-\langle \sigma(\psi,\Psi)\sigma(\psi,\Psi) \rangle/2}$.
In turn, at least formally, that identity determines the $n$-point functions (reproducing Wick's rule) by Stone theorem and (W2).
(b) From (CCR) it holds $\langle \sigma(\psi,\Psi)\sigma(\psi',\Psi)\rangle = \mu(\psi,\psi') - (i/2) \sigma(\psi,\psi')$,
where $\mu(\psi,\psi')$ is the symmetrized  two-point function 
$(1/2)(\langle \sigma(\psi,\Psi)\sigma(\psi',\Psi)\rangle + \langle  \sigma(\psi',\Psi)\sigma(\psi,\Psi)\rangle)$
 which defines a symmetric positive-semidefined bilinear form on $\sS$.
 (c) $\langle A^\dagger A\rangle\geq 0$
for  elements $A:= [e^{i\sigma(\psi,\Psi)} -I]+i[e^{i\sigma(\psi,\Psi)}-I]$ 
entails: 
\beq
|\sigma(\psi,\psi')|^2 \leq 4\:\mu(\psi,\psi)\mu(\psi',\psi')\:, \quad\quad \mbox{for every $\psi,\psi' \in \sS$}\label{sm}\:,
\eeq
which, in turn,  implies that {\em $\mu$ is strictly positive defined} because $\sigma$ is non degenerate. 
Reversing the procedure, the general definition of quasifree states on Weyl algebras is the following.\\

\definizione \label{defquasifree}
{\em Let $\cW_{\sS,\sigma}$ be  a Weyl algebra and $\mu$ a real scalar product on $\sS$ 
satisfying (\ref{sm}).
A state $\omega_\mu$ on $\cW_{\sS,\sigma}$ is called  {\bf quasifree state}  associated with $\mu$
if
\beq
\omega_\mu(W(\psi)) := e^{-\mu(\psi,\psi)/2} \:, \quad \mbox{for all $\psi\in \sS$.} \label{formal2}
\eeq} 
\lemma \label{lemma1}
{\em Let $(\sS,\sigma)$ be a real symplectic space with $\sigma$ non degenerate and $\mu$  
 a real scalar product on $\sS$ satisfying (\ref{sm}).
There is a complex Hilbert space $\cH_\mu$
and a map $K_\mu: \sS \to \cH_\mu$ with: 

(i)  $K_\mu$ is $\bR$-linear with dense complexified range, i.e. $\overline{K_\mu(\sS) + i K_\mu(\sS)}= \cH_\mu$,

(ii) for all $\psi,\psi' \in \sS$,
$\langle K_\mu\psi , K_\mu\psi'\rangle= \mu(\psi,\psi') - (i/2) \sigma(\psi,\psi')$.\\
Conversely, if the pair $(\cH,K)$ satisfies (i) and 
$\sigma(\psi,\psi')= -2 Im \langle K\psi , K\psi'\rangle_\cH$, with $\psi,\psi' \in \sS$, 
the unique real scalar product $\mu$ on  $\sS$ satisfying (ii) verifies  (\ref{sm}).} \\
  
\noindent The last statement arises by Cauchy-Schwarz inequality,  the remaining part being in Proposition 3.1 in \cite{KW}.
Notice that $K_\mu$ is always injective due to (ii) and non degenerateness of 
$\sigma$. Now existence of quasifree states can be proved using the lemma above with the following proposition. 
Therein, uniqueness and regularity of the state is contained in Lemma A.2 and Proposition 3.1 in \cite{KW}. \\

\proposizione\label{proposition2}
{\em For every $\mu$ as in definition \ref{defquasifree} the following hold.\\
{\bf (a)} there is a unique quasifree state $\omega_\mu$ 
associated with $\mu$ and it is  
regular.\\   
{\bf (b)} The GNS triple $(\gH_{\omega_\mu}, \Pi_{\omega_\mu}, \Upsilon_{\omega_\mu})$
   is determined as follows with respect to  $(\cH_\mu,K_\mu)$ in 
  lemma (\ref{lemma1}). (i) $\gH_{\omega_\mu}$ is the symmetric Fock space with  
  one-particle space $\cH_\mu$. (ii) The cyclic vector   $\Upsilon_{\omega_\mu}$ is the vacuum vector of $\gH_\omega$. 
   (iii) $\Pi_{\omega_\mu}$ is  determined by
$\Pi_{\omega_\mu}(W(\psi)) = e^{i\overline{\sigma(\psi,\Psi)}}$, the bar denoting the
closure, where\footnote{The field operator $\Phi(f)$, with $f$ in the complex Hilbert space $\gh$, 
used in \cite{BR2} in propositions 5.2.3 and 5.2.4 is related to $\sigma(\psi,\Psi)$ by means of
$\sigma(\psi,\Psi)= \sqrt{2} \Phi(iK_\mu \psi)$ assuming $\cH:= \gh$.}
\beq
\sigma(\psi,\Psi) :=  ia(K_\mu\psi) -ia^\dagger(K_\mu\psi)\:, \quad \mbox{for all $\psi\in\sS$} \label{fo}
\eeq
$a(\phi)$ and $a^\dagger(\phi)$, $\phi\in \cH_\mu$, being the usual annihilation (antilinear in $\phi$) 
and creation operators defined 
in the dense linear manifold spanned by the states with finite number of particles.\\
{\bf (c)} A pair $(\cH,K)\neq (\cH_\mu,K_\mu)$ satisfies (i) and (ii)  in lemma \ref{lemma1} 
for $\mu$, determining the same quasifree state $\omega_\mu$, if and only if there is a unitary operator
$U: \cH_\mu\to \cH$ such that $UK_\mu=K$.\\
{\bf (d)} $\omega_\mu$ is pure (i.e. its GNS representation is irreducible) 
if and only if $\overline{K_\mu(\sS)} = \cH_\mu$\footnote{In turn this is equivalent (see p.77 in \cite{KW}) to
$4\mu(\psi',\psi') =  \sup_{\psi\in \sS\setminus\{0\}} |\sigma(\psi,\psi')|/\mu(\psi,\psi)$ for every
$\psi'\in \sS$.}.}

\section{Spacetime infinities}\label{infinities}
From \cite{Wald} we give the following definition originally stated by Ashtekar \cite{AO}, see also \cite{AH} for fine distinctions on  requirements 
concerning  validity of vacuum Einstein equations.\\
 \definizione \label{defAF}{\em A time-oriented  four-dimensional  smooth spacetime $(M,g)$ satisfying vacuum Einstein equations is called 
{\bf vacuum spacetime asymptotically flat  at null and spatial infinity}, if 
there exists a spacetime $(\tM,\tg)$ with $\tg$ smooth everywhere except possibly a point $i^0$ (called {\bf spatial infinity}), 
where it is $C^{>0}$ (see p.227 of
\cite{Wald}), a diffeomorphism $\psi : M \to \psi(M) \subset \tM$  and a map $\Omega: \psi(M) \to [0,+\infty)$ so that
$\tg = \Omega^2 \psi^* g$ and the following facts hold. (We omit to write explicitly $\psi$ and $\psi^*$ in the following)\\
{\bf (1)} $\overline{J^+(i^0)} \cup \overline{J^-(i^0)} = \tM \setminus M$ the closure and causal sets being referred to $(\tM,\tg)$.
Thus $i^0$ is spacelike related with all the points of $M$ and the boundary $\partial M$ consists of the union of 
$\{i^0\}$, the {\bf future null infinity} $\scri = (\partial J^+(i^0))\setminus \{i^0\}$ and
and the {\bf past null infinity} $\scrip = (\partial J^-(i^0))\setminus \{i^0\}$.\\
{\bf (2)} There is a open neighborhood $V$ of $\partial M$ such that $(V,\tg)$ is strongly causal (see \cite{Wald}).\\
{\bf (3)} $\Omega$ can be extended to a function on $\tM$ which $C^2$ at least at $i^0$ and smooth elsewhere.\\
{\bf (4)} (a) $\Omega\spa \rest_{\scri \cup \scrip} =0$ but $d\Omega(x) \neq 0$ for $x\in \scri \cup \scrip$.
(b) $\Omega(i^0) =0$ and the limits toward $i^0$ of $d\Omega$ and $\tilde{\nabla}_\mu \tilde{\nabla}_\nu \Omega$
are respectively $0$ and $2\tg_{\mu\nu}(i^0)$.\\
{\bf (5)} (a) The map of null directions at $i^0$ into the space of integral curves of $n^\mu:= \tilde{\nabla}^\mu \Omega$ on $\scri$
and $\scrip$ is a diffeomorphism. (b) For a strictly positive smooth function $\omega$ defined in a neighborhood 
of $\scri \cup \scrip$ which satisfies $\tilde{\nabla}_\mu (\omega^4 n^\mu) =0$ on $\scri \cup \scrip$
the integral curves of $\omega^{-1}n$ are complete on $\scri \cup \scrip$.}

\section{Fourier-Plancherel transform on $\bR\times \bS^2$.} \label{Fourier}
 Define $\cS(\scri;\bC):= \cS(\scri)+ i \cS(\scri)$ 
(i.e. the complex linear space of the complex-valued smooth functions $\psi:\scri \to \bR$ such that,
in that Bondi frame,
$\psi$ with all derivatives vanish as $|u|\to +\infty$, uniformly in $\z,\bz$,
faster than $|u|^{-k}$, $\forall k\in \bN$).  The space
$\cS(\scri;\bC)$ 
generalizes straightforwardly Schwartz' space on $\bR^n$. 
It can be equipped with the Hausdorff topology induced from the countable class seminorms --
{\em they depend on the Bondi frame but the topology does not} -- $p,q,m,n \in \bN$,
$$||\psi||_{p,q,m,n} := \sup_{(u,\z,\bz)\in \scri} \left||u|^p \partial^q_u\partial^m_{\z}\partial^n_{\bz}\psi(u,\z,\bz)\right|
\:.$$
$\cS(\scri;\bC)$ is dense in both $L^1(\bR\times \bS^2, du\wedge \epsilon_{\bS^2}(\z,\bz))$ and 
$L^2(\bR\times \bS^2, du\wedge \epsilon_{\bS^2}(\z,\bz))$ (with the topology of these spaces
which are weaker than that of $\cS(\scri;\bC)$),
 because it includes the dense space 
%%%%CORREZIONE cambiato C_c in C_0
$C_0^\infty(\bR\times \bS^2; \bC)$
%%%
of smooth compactly supported complex-valued functions. We also define the space of {\em distributions}
$\cS'(\scri;\bC)$ containing all the linear functionals from $\bR\times \bS^2$ to $\bC$ 
which are weakly continuous with respect to the topology of  $\cS(\scri;\bC)$. Obviously
$\cS(\scri;\bC)\subset \cS'(\scri;\bC)$ and $L^p(\bR\times \bS^2, du\wedge \epsilon_{\bS^2}(\z,\bz))
\subset \cS'(\scri;\bC)$ for $p=1,2$.
We introduce the Fourier transforms $\cF_\pm(\psi)$ of $f\in \cS(\scri;\bC)$
$$\cF_\pm(\psi)(k,\z,\bz) := \int_{\bR\times \bS^2} \frac{e^{\pm i ku}}{\sqrt{2\pi}} f(u,\z,\bz) 
du \wedge \epsilon_{\bS^2}(\z,\bz)\:, 
\quad (k,\z,\bz)\in \bR \times \bS^2\:.$$ 
$\cF_\pm$  the properties listed within the theorem below whose proof is a straightforward extensions of the analog
for standard Fourier transform in $\bR^n$
(theorems IX.1, IX.2, IX.6, IX.7 in \cite{RS}). In (4) $C_\infty(\scri)$ denotes the Banch space,
 with respect to the supremum norm $||\cdot||_\infty$, of the continuous complex valued functions on $\bR\times \bS^2$ 
 {\em vanishing at infinity}, i.e. $f\in C_\infty(\bR\times \bS^2)$ iff $f$ is continuous and, for every $\epsilon>0$
 there is a compact set $K_\epsilon \subset \bR\times \bS^2$ with $|f(x)|< \epsilon$ for $x\not \in K_\epsilon $.\\

\teorema \label{fourier}{\em The maps $\cF_\pm$ satisfy the following properties.\\
{\bf (1)} for all $p,m,n \in \bN$ and every $\psi \in \cS(\scri;\bC)$ it holds 
$$\cF_\pm\left(\partial^p_u \partial^m_\z \partial^n_{\bz} \psi\right)(k,\z,\bz) = (\pm i)^p k^p \partial^m_\z \partial^n_{\bz} 
\psi 
\cF_\pm(\psi)(k,\z,\bz)\:.$$
{\bf (2)} $\cF_\pm$ are continuous bijections onto $\cS(\scri;\bC)$ and $\cF_-= (\cF_+)^{-1}$.\\
{\bf (3)} If $\psi,\phi \in \cS(\scri;\bC)$ one has
\begin{eqnarray} \int_\bR \overline{\cF_\pm(\psi)}(k,\z,\bz)\cF_\pm(\phi)(k,\z,\bz) dk &=& 
\int_\bR \overline{\psi}(u,\z,\bz)\phi(u,\z,\bz) du \:,\: \mbox{for all $(\z,\bz) \in \bS^2$}\:,\nonumber\\
\int_{\bR\times \bS^2} \overline{\cF_\pm(\psi)(k,\z,\bz)}\cF_\pm(\phi)(k,\z,\bz) dk\wedge \epsilon_{\bS^2}(\z,\bz) &=& 
\int_{\bR\times \bS^2} \overline{\psi(u,\z,\bz)}\phi(u,\z,\bz) du\wedge \epsilon_{\bS^2}(\z,\bz)\:.\nonumber
\end{eqnarray}
{\bf (4)} If $T\in \cS'(\scri;\bC)$ the definition
$\cF_\pm{T}(f):= T\left(\cF_\pm(f)\right)\:, 
 \mbox{for all
 $f\in \cS'(\scri;\bC)$}$
 is well-posed, gives rise to  the unique weakly continuous linear extension of 
 $\cF_\pm$ to $\cS'(\scri;\bC)$ and one has, with the usual definition of derivative of a distribution,
 $$\cF_\pm \left(\partial^p_u \partial^m_\z \partial^n_{\bz} T\right) = 
 (\pm i)^p k^p \partial^m_\z \partial^n_{\bz}\cF_\pm(T)\:, \quad \mbox{for all $p,m,n \in \bN$}\:.$$
{\bf (5) Plancherel theorem}. From (3) and reminding that $\cS(\scri;\bC)$ is dense in 
  the Hilbert space $L^2(\bR\times \bS^2, du\wedge \epsilon_{\bS^2}(\z,\bz))$,
  $\cF_\pm$ extend uniquely to unitary transformations 
   from the Hilbert space 
  $L^2(\bR\times \bS^2, du\wedge \epsilon_{\bS^2}(\z,\bz))$ to $L^2(\bR\times \bS^2, du\wedge \epsilon_{\bS^2}(\z,\bz))$
 and the extension of $\cF_-$ is the inverse of that of $\cF_+$.\\
 These extensions coincide respectively with the restrictions
 to $L^2(\bR\times \bS^2, du\wedge \epsilon_{\bS^2}(\z,\bz))$ of the action of $\cF_\pm$ on distributions as in (4)
 when reminding that $L^2(\bR\times \bS^2, du\wedge \epsilon_{\bS^2}(\z,\bz))
\subset \cS'(\scri;\bC)$.\\
{\bf (6) Riemann-Lebesgue lemma}. 
 Reminding  that $\cS(\scri;\bC)$ is dense in  
  $L^1(\bR\times \bS^2, du\wedge \epsilon_{\bS^2}(\z,\bz))$, $\cF_\pm$ uniquely extend
 to a bounded operator from $L^1(\bR\times \bS^2, du\wedge \epsilon_{\bS^2}(\z,\bz))$ to 
 $C_\infty(\bR\times \bS^2)$. In particular one has, for $f\in L^1(\bR\times \bS^2, du\wedge \epsilon_{\bS^2}(\z,\bz))$
 $$||\cF_\pm (f) ||_\infty \leq (2\pi)^{-1/2} ||f||_1$$
These extensions coincide respectively with the restrictions
 to $L^1(\bR\times \bS^2, du\wedge \epsilon_{\bS^2}(\z,\bz))$ of the action of $\cF_\pm$ on distributions as in (4)
 when reminding that $L^1(\bR\times \bS^2, du\wedge \epsilon_{\bS^2}(\z,\bz))
\subset \cS'(\scri;\bC)$.}\\

\noindent From now on $\cF : \cS'(\scri;\bC) \to \cS'(\scri;\bC)$ denotes the extension to distributions of $\cF_+$
as stated in (4) in theorem \ref{fourier} whose inverse, $\cF^{-1}$, is the analogous extension of $\cF_-$. 
We call $\cF$ {\bf Fourier-Plancherel transformation}, also if, properly speaking this name should be reserved to its 
restriction to $L^2(\bR\times \bS^2, du\wedge \epsilon_{\bS^2}(\z,\bz))$ defined in (5) in theorem \ref{fourier}.
We also use the formal distributional notation for $\cF$ (and the analog for $\cF^{-1}$) 
$$\cF(\psi)(k,\z,\bz) := \int_{\bR\times \bS^2} \frac{e^{i ku}}{\sqrt{2\pi}} f(u,\z,\bz) du \wedge \epsilon_{\bS^2}(\z,\bz)\:, $$ 
 regardless if $f$ is  a function or a distribution. We have the following final proposition whose proof
is immediate from (4) and (5) in theorem \ref{fourier}.\\

\proposizione \label{last} {\em Let $m\in \bN$. The Fourier-Plancherel transform $\cF(T)$ of a distribution $T\in \cS'(\scri;\bC)$ 
is a measurable function satisfying 
$$\int_{\bR\times \bS^2} (1+ |k|^2)^m |\cF(T)|^2 dk\wedge \epsilon_{\bS^2}(\z,\bz)<+\infty$$
if and only if the $u$-derivatives of $T$
 in the sense of distributions, are measurable functions and satisfy 
$$\partial^n_u T \in L^2(\bR\times \bS^2, du\wedge \epsilon_{\bS^2})\:,\:\: \mbox{for $\bN \ni n\leq m$}. $$}\\

\end{document}